\documentclass{iopjournal_arxiv}
\usepackage[OT4]{fontenc}
\usepackage{siunitx}
\DeclareSIUnit\dBc{dBc} 

\DeclareSIUnit\MegaSamples{MS} 
\usepackage{ragged2e} 
\usepackage{graphicx}
\usepackage{multirow} 
\usepackage{amssymb}
\usepackage{threeparttable}

\begin{document}
\justifying

\articletype{Paper} 

\title{Unlimited-Range Absolute Distance Interferometry Outdoors via Tunable Synthetic and Pseudo-Supersynthetic Wavelengths: The TeleYAG-II Interferometer}

\author{Florian Pollinger$^{1,*}$\orcid{0000-0002-8494-4687}, Tobias Meyer$^1$\orcid{0000-0002-7991-0837}, Frank Pilarski$^1$ \orcid{0009-0006-7669-6931},  Hongdan Yan$^1$\orcid{0000-0002-3529-5596}, Anni Sauthoff$^1$\orcid{0000-0002-5954-9677},  G\"{u}nther Prellinger$^1$\orcid{0000-0003-4919-0258},  Stephanie Weinrich$^1$\orcid{0009-0004-8169-0847}, Frank Schmaljohann$^1$\orcid{0000-0002-8739-8458} and Paul K\"{o}chert$^{1}$\orcid{0000-0002-2583-6643}}

\affil{$^1$Physikalisch-Technische Bundesanstalt (PTB), Bundesallee 100, 38116 Braunschweig, Germany}

\affil{$^*$Author to whom any correspondence should be addressed.}

\email{florian.pollinger@ptb.de}


\begin{abstract}
\justifying
Absolute distance interferometry requires an unlimited non-ambiguity range to be truly absolute. Tunable synthetic wavelength interferometry allows for the construction of pseudo-supersynthetic wavelengths using a limited number of optical sources and a standard interferometer setup. This approach facilitates the creation of a series of consecutively deployable synthetic wavelengths, ultimately combining macroscopic unambiguity with microscopic interferometric accuracy. We have implemented and demonstrated this concept in the TeleYAG-II interferometer, designed as an SI-traceable standard for the calibration of geodetic baselines. The system employs two Nd:YAG lasers as optical sources, with their frequency difference controlled by a versatile optical phase-locked loop. Specifically engineered for robustness in field conditions, the heterodyne interferometer was validated through an outdoor measurement campaign. The results demonstrate the efficacy of the method, achieving an expanded measurement uncertainty of 0.13 mm over a distance of 224 m without the need for auxiliary length measurements.
\end{abstract}

\section{Introduction} \label{Sec_Introduction}
Geodesy and surveying are based on long distance measurements of high accuracy. From everyday cadastral work \cite[chap.~VI]{unece1996land}, over hazard and deformation monitoring \cite[chap.~13]{Niemeier2008}  to the maintenance of geodetic reference frames \cite[p.~242]{Gross2009}, long distance measurements, both optical or GNSS-based are an essential element of these disciplines. Providing traceability and verification for such measurements has always been a challenge for dimensional metrology \cite{Baird1968, Rueeger1996}. A technology-open verification strategy is the use of geodetic baselines with known reference distances of high accuracy \cite{Pollinger2012,  Wu2017, Wisniewski2026, Jokela2026}. To establish the reference distances with sufficient accuracy, however, suitable reference standards are required. There are two fundamental challenges to be overcome. Firstly, how can long-range capability and high accuracy be combined in one instrument? Secondly, how can such an instrument be constructed to withstand the challenges of outdoor measurements, i.e. be truly field-capable?

In 1923, V\"{a}is\"{a}l\"{a} suggested the use of white light interferometry as tool to realise long distances by optical multiplication of a reference standard of high length accuracy \cite{Vaisala1923}. The Finnish Geospatial Research Institute successfully applies the  V\"{a}is\"{a}l\"{a} White Light Interferometer for the calibration of their Nummela baseline up to today, achieving sub-millimetre accuracies for distances up to \qty{864}{\metre} \cite{Jokela2026}. But the method has some serious drawbacks. First, it requires absolutely stable environmental conditions, which means that campaigns last for several weeks and are therefore economically less viable. Furthermore, it can only be applied to dedicated baselines as it can only measure multiples of the initial reference length. Baselga et al. developed a rigorously SI-traceable distance measurement based on global navigation satellite systems (GNSS) in range up to \qty{5}{\kilo\metre}. While this technology does not require line-of-sight and does not depend on an auxiliary length pre-value,  observation times over ten hours are considered minimum \cite{Baselga2022}. Exploiting the progress of photonics and electronics in the last decades, a number of different optical measurement approaches have been demonstrated in recent years. Maybe the most advanced approach with respect to accuracy, applicability and field capability, Guillory et al. have recently pushed the classic concept of radio-frequency (RF) modulation to new levels of accuracy \cite{Guillory2016, Guillory2019, Guillory2023}.   But also other optical, maybe less established measurement principles have proven field capability in the last years. Systems based on the inter-mode beat of femtosecond lasers  \cite{Minoshima2000, Pollinger2012, Xu2020} or dual-comb asynchronous optical pulse sampling \cite{Kim2023, Jiang2025} have successfully been deployed outdoors. Wu et al. demonstrated interferometric absolute distance measurements up to \qty{1.2}{\kilo\meter}  using dual-comb multi-heterodyne interferometry based on dual electro-optic combs \cite{Wu2017}. With the TeleYAG interferometer, our group demonstrated synthetic wavelength interferometry over distances up to \qty{864}{\metre} \cite{MeinersHagen2015, MeinersHagen2017}. All the more modern optical methods, however, require a coarse prior knowledge of the distance. The aim of this work is to overcome the ambiguities inherent in multi-wavelength interferometry (MWLI). For this purpose, we propose a new MWLI approach: the combination of interferometry with tunable synthetic wavelength and interferometry with pseudo-supersynthetic wavelength. It is introduced and elaborated in Sec.~\ref{Sec_Method}. It is implemented in the new TeleYAG-II interferometer, the successor of the TeleYAG instrument. The latter demonstrated the applicability of heterodyne interferometry to the optical long-distance measurement problem. However, the original instrument had serious flaws which prevented its use as a primary standard. The alignment of optical source and interferometer head was highly unstable under field conditions, requiring frequent realignments during the measurement. The data had to be heavily filtered for outliers. There was no fixed physical reference point, and above all, the pre-value of the length had to be accurate to the millimetre level for the correct unwrapping. As part of the European metrology research project 'Geometre' \cite{Pollinger2023} the lessons learned from the TeleYAG instrument were taken to develop a more-field capable distance meter.  Great care has been put in the instrumental design of source and interferometer head. It is discussed in detail in Section \ref{Sec_Experimental}. Typical data characterising the instrument and elaborating the method is compiled in Sec.~\ref{Sec_Results}. A measurement uncertainty budget is derived in Sec.~\ref{Sec_MU}, followed by a more general discussion and conclusions in Sec.~\ref{Sec_Discussion}.

\section{Method} \label{Sec_Method}
In this section, we develop the concept of pseudo-supersynthetic wavelength interferometry for practical absolute distance interferometry. As basis, the established concept of iterative unwrapping in fixed synthetic wavelength interferometry is recapitulated in section \ref{Sec_Method_SWL}, before being expanded to the new concepts of tunable synthetic wavelength interferometry in section \ref{Sec_Meth_OPLL_and_Variable_Fixed_Synthetic_Wavelength} and finally of pseudo-supersynthetic wavelength interferometry in section \ref{Sec_Method_PSSL}.

\subsection{Synthetic Wavelength Interferometry} \label{Sec_Method_SWL}
 For a given geometric distance $l$ and a phase index of refraction $n^{\rm ph}(\lambda)$, the interferometric optical phase $\phi( l, \lambda, t)$ measured with the optical vacuum wavelength $\lambda$ at the time $t$ is given by  
\begin{equation}
\phi(l,t, \lambda)=4\pi \frac{n^{\rm ph}(\lambda) l(t)}{\lambda(t)} \label{Eq_Interferometric_Phase_basic}
\end{equation}
In principle, the optical phase $\phi$ can only be resolved modulo $2 \pi$, limiting the  non-ambiguity range (NAR) of interferometric measurements to half the optical wavelength $\lambda/2$. In classical interferometry, the NAR is extended by moving the reflector along the distance to be measured and counting the number $N$of full $2 \pi$ completions (or fringes) along the path. For the calibration of baselines, however, distances of several hundred up to several thousand metres are to be measured \cite{Pollinger2012, Wisniewski2026}. A continuous measurement is practically impossible under field conditions. A true absolute distance measurement, i.e. a single measurement from start to end point is necessary. How to unwrap this (highly accurate) measurement and thus, to combine macroscopic relevance with microscopic accuracy is the holy grail in absolute distance interferometry. Several approaches have been developed \cite{MeinersHagen2009, Falaggis2018}. The idea to combine the phase information of multiple lasers to enlarge the NAR had already been proposed by Beno\^{\i}t \cite{Benoit1898} very shortly after the first groundbreaking papers by Michelson \cite{Michelson1894}. Synthetic wavelength (SWL) interferometry, a MWLI variant \cite{Wyant1971,  MeinersHagen2009, Falaggis2018}  is based on the idea to combine the phase information $\phi_1(t)$ and $\phi_1(t)$ of interferometer measurements with different wavelengths $\lambda_1 >  \lambda_2$ at the same time:
\begin{equation}
\phi_{1,2}^{\rm swl}  (l,t)=\phi_2 (l,t,\lambda_2 )-\phi_1 (l,t,\lambda_1 ). \label{Eq_SynPhase}
\end{equation}
Inserting Eq.~(\ref{Eq_Interferometric_Phase_basic}), one can easily derive
\begin{equation}
\phi_{1,2}^{\mathrm{swl}}(l,t)=4\pi n^{\rm 2WL}_{1,2}\frac{l(t)}{ \Lambda_{1,2}^{\mathrm{swl}}}, \label{Eq_Length_From_Synthetic_WL}
\end{equation}
introducing the two-colour group index of refraction 
\begin{equation}
n^{\rm 2WL}_{1,2} =n^{\rm ph}(\lambda_1)-\frac{n^{\rm ph}(\lambda_1)-n^{\rm ph}(\lambda_2)}{\lambda_1-\lambda_2}\lambda_1. \label{Eq_Two_colour_index_of_refraction}
\end{equation}
and the synthetic wavelength
\begin{equation}
\Lambda_{1,2}^{\mathrm{swl}} = \frac{\lambda_1\lambda_2}{\lambda_1-\lambda_2}.  \label{Eq_Def_Synthetic_Wavelength_by_wavelengths}
\end{equation}
Half the synthetic wavelength $\Lambda_{1,2}^{\mathrm{swl}}/2$ defines the magnitude of the NAR of the combined measurement. For specific, highly stable lasers and environments, such as gauge block interferometers,  de Groot showed in 1994 that  this NAR can be extended by a few multiple orders of  $\Lambda_{1,2}^{\mathrm{swl}}/2$ \cite{Groot1994}, but in general cases, the NAR is limited to exactly $\Lambda_{1,2}^{\mathrm{swl}}/2$. From this point of view, a long synthetic might seem favourable. However, as a consequence of Eq.~(\ref{Eq_Length_From_Synthetic_WL}), the uncertainty $u(l_{1,2})$ of the distance measurement scales proportional with the synthetic wavelength $\Lambda_{1,2}^{\mathrm{swl}}$. Simply put, the longer the NAR, the bigger the uncertainty of the measurement result. To combine a macroscopic NAR with a microscopic accuracy, it is an established strategy in MWLI to measure a geometric distance $l$ simultaneously with $n$ different laser wavelengths $\lambda_1, \lambda_2, \lambda_3, ...$. A synthetic phase can be calculated for each pairwise combination of the measured optical phases $\left(\phi_i(l, t, \lambda_i), \phi_j(l, t, \lambda_j)\right)$ with $i \neq j$ and $i,j = 1, 2, ..., n$ by Eq.~(\ref{Eq_SynPhase}). These synthetic phases can be sorted in series 
\begin{equation}
\phi_{i_1,j_1}^{\rm swl}(d,t, \lambda_{i_1}, \lambda_{j_1}),\ \phi_{i_{2},j_{2}}^{\rm swl}(d,t, \lambda_{i_2}, \lambda_{j_2}),\ \phi_{i_{3},j_{3}}^{\rm swl}(d,t, \lambda_{i_3}, \lambda_{j_3}),\ ... \label{Eq_Ordered_Synthetic_Phases_Series}
\end{equation}
 for which the corresponding synthetic wavelengths according to Eq.~(\ref{Eq_Def_Synthetic_Wavelength_by_wavelengths}) 
\begin{equation}
\Lambda_{i_1,j_1}^{\rm swl},\ \Lambda_{i_{2},j_{2}}^{\rm swl},\ \Lambda_{i_{3},j_{3}}^{\rm swl},\ ... \label{Eq_Ordered_Synthetic_Wavelengths_Series}
\end{equation}
 fulfill
\begin{equation}
\Lambda_{i_1,j_1}^{\rm swl} > \Lambda_{i_{2},j_{2}}^{\rm swl} >  \Lambda_{i_{3},j_{3}}^{\rm swl} >\ ... \label{Eq_Condition_Series_Synthetic_Wavelengths_By_Lengths}
\end{equation}
and the corresponding uncertainties $u$ of the distances $l_{i_1, j_1}, \  l_{i_2, j_2},\  l_{i_3, j_3},\ ...$ derived according to (\ref{Eq_Length_From_Synthetic_WL}) adhere
\begin{equation}
u \left( l_{i_{k-1}, j_{k-1}} \right) < \frac{\Lambda_{i_{k},j_k}^{\rm swl}}{4}.  \label{Eq_Condition_Series_uncertainty_lambda_fourth}
\end{equation}
These measurement series can be used to iteratively unwrap more accurate measurements with coarser measurement results of larger unambiguity using \cite{Pollinger2009}
\begin{eqnarray}
N_k 	& = & {\rm floor} \left( \frac{  2 l_{i_{k-1}, j_{k-1}}}{\Lambda_{i_{k},j_k}^{\rm swl}} - \frac{\phi_{i_{k},j_k}^{\mathrm{swl}}}{2 \pi n^{\rm 2WL}_{i_k, j_k}}+\frac{1}{2} \right) \label{Eq_Unwrap_Order_Number} \\
l_k 	& = & \frac{\Lambda_{i_{k},j_k}^{\rm swl}}{2} \left( N_k + \frac{\phi_{i_{k},j_k}^{\mathrm{swl}}}{2 \pi n^{\rm 2WL}_{i_k, j_k}} \right)\label{Eq_LengthFrom2Colour}
\end{eqnarray}
with the ${\rm floor}(x)$ function returning the largest integer value $z$ with $z<x$.

\subsection{OPLL-Mediated Tunable Synthetic Wavelength Interferometry} \label{Sec_Meth_OPLL_and_Variable_Fixed_Synthetic_Wavelength}
\begin{figure}
 \centering
	\includegraphics[width =1\columnwidth]{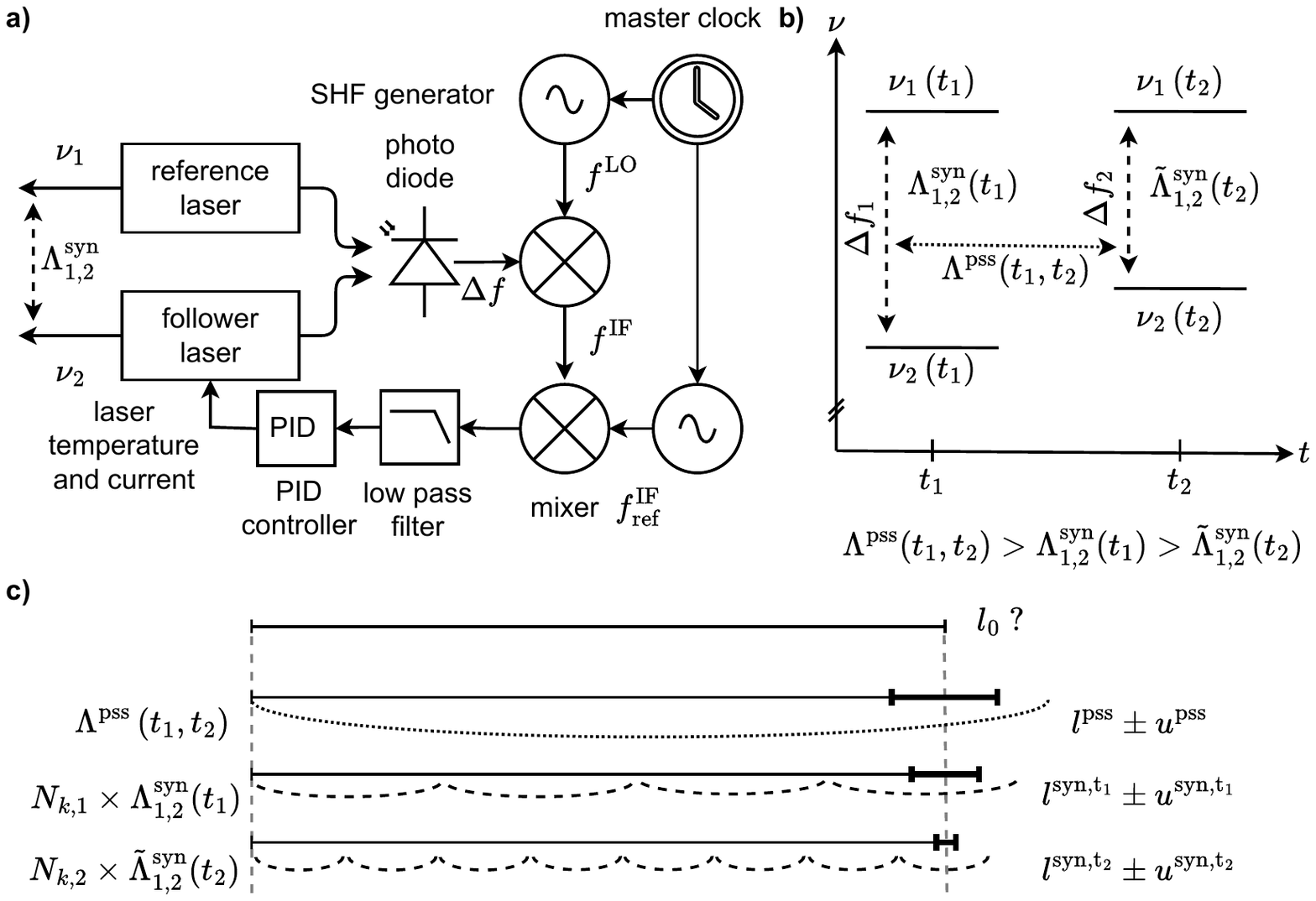} 
 \caption{a) Schematic design of the optical phase-locked loop (OPLL). b) Generation of the tunable synthetic wavelengths $\Lambda^{\rm swl}_{1,2}(t_1)$ and $\tilde{\Lambda}^{\rm swl}_{1,2}(t_2)$ by variation of the frequency difference $\Delta f$ and computation of their mutual pseudo-supersynthetic wavelength $\Lambda^{\rm pss}_{1,2}(t_1, t_2)$ indicated in the time-frequency diagram. c) Consecutive unwrapping using the series of asynchronous synthetic and pseudo-supersynthetic wavelengths indicated in b).}
\label{Fig_concept_PSS}
\end{figure}
A literal experimental realisation of such a wavelength series defined by equations (\ref{Eq_Condition_Series_Synthetic_Wavelengths_By_Lengths}) and (\ref{Eq_Condition_Series_uncertainty_lambda_fourth}) requires a complex laser system. Examples from the literature are based on CO$_2$ lasers \cite{Bourdet1979, Matsumoto1984}, electro-optic combs \cite{Yang2014, Yang2015, Zhao2018, Park2024, Li2025}, or fibre combs \cite{Yokoyama2009, Berg2012,  Zhu2018}. Interferometry using tunable laser sources allows for a NAR that is, in principle, unlimited. Working with tunable wavelength sources is sometimes used to measure the distance absolutely \cite{Stone1999, Kinder2002, Dale2014, Prellinger2015} (often referred to as frequency sweeping interferometry (FSI) or variable synthetic wavelength interferometry (VSW)). The 'synthetic wavelength' in this case, the bandwidth of the frequency variation, determines and often limits the  uncertainty of the distance measurement. They are also notoriously sensitive to instabilities in the measurement setup during the frequency variation. To enhance the accuracy of these absolute measurements, FSI values have sometimes been  used as coarse pre-values for SWL measurements  \cite{Pollinger2009, Hartmann2008, Floch2008,  Zhang2018}.

Absolute distance interferometry over several hundreds of metres defines strict requirements on the temporal coherency of the laser source, requiring a line width below \qty{150}{\kilo\hertz} for a distance of \qty{1}{\kilo\metre}.
Diode--pumped solid state lasers like Nd:YAG lasers fulfil this requirement. Their finite tuning lasing bandwidth of approximately \qty{30}{\giga\hertz} allows the formation of synthetic wavelengths. As shown in Fig.~\ref{Fig_concept_PSS}a, a synthetic wavelength can be generated by stabilising the frequency difference 
\begin{equation}
\Delta f  = \nu_2 - \nu_1 \label{Eq_Definition_Delta_f_lasers} 
\end{equation}
between two separate laser resonators using an optical phase-locked loop (OPLL) \cite{MeinersHagen2015, MeinersHagen2017, Gelmini1994, Gelmini1995, Azouigui2010}. Inserting $c = \lambda_i \nu_i (i = 1,2)$ to Eq.~(\ref{Eq_Def_Synthetic_Wavelength_by_wavelengths}) (with $c$ representing the vacuum speed of light),  the synthetic wavelength between the both lasers can be expressed as \cite{MeinersHagen2017}
\begin{equation}
\Lambda_{1,2}^{\mathrm{swl}} =\frac{c}{\Delta f}\label{Eq_Def_Synthetic_Wavelength_by_frequency}.
\end{equation}
Gelmini et al. pointed out in the 1990s that Eq.~(\ref{Eq_Def_Synthetic_Wavelength_by_frequency}) opens the possibility to construct further synthetic wavelengths with a laser pair whose frequency difference can be controlled \cite{Gelmini1994, Gelmini1995}. In principle, the OPLL frequency can be freely selected within the active laser medium of both resonators and must be detectable by a suitable photodiode circuit. If  the synthetic multiples described  in Sec. \ref{Sec_Method_SWL} can be measured in succession, it is possible to generate series of measurements with different synthetic wavelengths that satisfy conditions (\ref{Eq_Condition_Series_Synthetic_Wavelengths_By_Lengths}) and (\ref{Eq_Condition_Series_uncertainty_lambda_fourth}) by simply varying the frequency $f_{\rm LO}$ of the super-high-frequency (SHF) generator (see Fig.~\ref{Fig_concept_PSS}). Thus, the OPLL stabilisation between the two Nd:YAG lasers acts as an efficient multi-wavelength generator and can be used directly for tunable synthetic wavelength interferometry (TSWL) for the purposes of this work. 

\subsection{Pseudo-Supersynthetic Wavelength Interferometry} \label{Sec_Method_PSSL}
The multi-wavelength generator described in the previous section \ref{Sec_Meth_OPLL_and_Variable_Fixed_Synthetic_Wavelength} offers the possibility to generate synthetic wavelengths within a range limited at the shorter end by the gain bandwidth of both laser resonators (or, if this is smaller, by the bandwidth of the photodetector) and at the longer end by the intermediate frequency $f^{\rm IF}$. However, there are practical limitations to the full realisation of a tuning range from, for example, \qty{1}{\mega\hertz} (\qty{600}{\metre}) to \qty{20}{\giga\hertz} (\qty{30}{\milli\metre}). For once, carrying out the measurements sequentially for all these values takes a considerable amount of time. Furthermore, in practice, Nd:YAG laser sources cannot be tuned completely continuously across the entire gain bandwidth due to unavoidable mode hopping (a typical mode hop free spectral range here: \qty{5.6}{\giga\hertz}). This further complicates and delays the measurement, meaning that a complete realisation of the synthetic wavelength series using this approach is not feasible in practice.

Fortunately, there is a way to further increase the NAR by taking into account phase measurements taken at different  times $t_1$ and $t_2$  using two synthetic signals with high spatial resolution ($\Lambda^{\rm swl}_{i_1, j_1}$ and $\Lambda^{\rm swl}_{i_2, j_2}$). This allows two measurements with a NAR in the millimetre range to be combined in such a way that the unambiguously determinable measurement range can be extended to the metre range (without leaving the mode hop free region of the resonators). Assume two synthetic phases $\phi^{\rm swl, 1}_{1,2} \left( t_1, \Delta f_1 \right)$ and $\phi^{\rm swl, 2}_{1,2} \left( t_2, \Delta f_2 \right) $ with $\Delta f_2 > \Delta f_1$ determined from 
\begin{eqnarray}
\phi^{\rm swl, 1}_{1,2} \left( t_1, \Delta f_1 \right) 	& = &	\phi_1\left(t_1 \right) - \phi_2\left(t_1 \right) \nonumber \\
\phi^{\rm swl, 2}_{1,2} \left( t_2, \Delta f_2 \right) 	& = &	\phi_1\left(t_2 \right) - \phi_2\left(t_2 \right) \label{Eq_Synthetic_phases_for_PSS}
\end{eqnarray}
In complete analogy to Eq.~(\ref{Eq_SynPhase}), they can be subtracted from each other, resulting in a phase difference:
\begin{equation}
\phi^{\rm pss}(t_1, \Delta f_1, t_2, \Delta f_2) = \phi^{\rm swl, 1}_{1,2} \left( t_1, \Delta f_1 \right)  - \phi^{\rm swl, 2}_{1,2} \left( t_2, \Delta f_2 \right). \label{Eq_Def_PSS_Phase} 
\end{equation}
Due to the similarity to the synthetic phase, we propose the term pseudo-supersynthetic (PSS) phase. Further to this, applying the synthetic wavelength model according to Eq.~(\ref{Eq_Def_Synthetic_Wavelength_by_wavelengths}) and (\ref{Eq_Def_Synthetic_Wavelength_by_frequency}), the pseudo-supersynthetic wavelength
\begin{eqnarray}
\Lambda^{\rm pss} (t_1, \Delta f_1, t_2, \Delta f_2)  &=& \frac{\Lambda_{1,2}^{\mathrm{swl}}(t_1, \Delta f_1) \Lambda_{1,2}^{\mathrm{swl}}(t_2, \Delta f_2)}{\Lambda_{1,2}^{\mathrm{swl}}(t_1, \Delta f_1)-\Lambda_{1,2}^{\mathrm{swl}}(t_2, \Delta f_2)} \label{Eq_Def_PSS_Wavelength_From_SynWLs} \\
 &=& \frac{c}{\Delta f_2 - \Delta f_1} \label{Eq_Def_PSS_Wavelength_From_Delta_fs}
\end{eqnarray}
can be defined. The geometric length $l^{\rm pss}$ can be derived from known PSS phase and wavelength using
\begin{equation}
\phi^{\rm pss}(t_1, \Delta f_1, t_2, \Delta f_2)=4\pi n^{\rm pss}(t_1, t_2)\frac{l^{\rm pss}}{ \Lambda^{\rm pss} (t_1, \Delta f_1, t_2, \Delta f_2) }, \label{Eq_Length_From_PSS_Measurement}
\end{equation}
with the PSS index of refraction 
\begin{equation}
n^{\rm pss}(\Lambda_1^{\rm swl}, \Lambda_2^{\rm swl}) =n^{\rm 2WL}_{1,2} (\Lambda_1^{\rm swl})-\frac{n^{\rm 2WL}(\Lambda_1^{\rm swl})-n^{\rm 2WL}(\Lambda_2^{\rm swl})}{\Lambda_1^{\rm swl}-\Lambda_2^{\rm swl}}\Lambda_1^{\rm swl}. \label{Eq_PSS_index_of_refraction}
\end{equation}
in complete formal analogy to Eq.~(\ref{Eq_Length_From_Synthetic_WL}) and (\ref{Eq_Two_colour_index_of_refraction}).

The benefit of this formalism can be understood from Eq.~(\ref{Eq_Def_PSS_Wavelength_From_Delta_fs}):  by taking measurements using successive synthetic signals whose frequency differences $\Delta f_1$ and $\Delta f_2$ are slightly out of tune with one another but  not identical, it is possible to construct very long PSS wavelengths $\Lambda_{1}^{\rm pss}, \Lambda_{2}^{\rm pss}, ...$ (inversely proportional to $\Delta f_{i, \rm pss} = \Delta f_{i,2} - \Delta f_{i,1}$), and thus the wavelength series from Eq.~(\ref{Eq_Ordered_Synthetic_Phases_Series}) can be extended to include longer NARs by adding PSS wavelengths

\begin{equation}
\Lambda_{1}^{\rm pss}, \Lambda_{2}^{\rm pss}, ... \Lambda_{i_1,j_1}^{\rm swl},\ \Lambda_{i_{2},j_{2}}^{\rm swl},\ \Lambda_{i_{3},j_{3}}^{\rm swl},\ ... \label{Eq_Ordered_PSS_and_Synthetic_Wavelengths_Series}
\end{equation}
 with
\begin{equation}
\Lambda_{1}^{\rm pss} > \Lambda_{2}^{\rm pss} > ... >\Lambda_{i_1,j_1}^{\rm swl} > \Lambda_{i_{2},j_{2}}^{\rm swl} >  \Lambda_{i_{3},j_{3}}^{\rm swl} >\ ... \label{Eq_Condition_Series_PSS_and_Synthetic_Wavelengths_By_Lengths}
\end{equation}
and 
\begin{equation}
u \left( l_{i_{k-1}, j_{k-1}} \right) < \frac{\Lambda_{i_{k},j_k}^{\rm swl/pss}}{4}.  \label{Eq_Condition_PSS_Syn_Series_uncertainty_lambda_fourth}
\end{equation}
Figures~\ref{Fig_concept_PSS}b and c illustrate how such a series of three wavelengths (one PSS, two SWLs) can be constructed from a consecutive measurement (Fig.~\ref{Fig_concept_PSS}b) and how a geometric length can subsequently be determined using an iterative algorithm (Fig.~\ref{Fig_concept_PSS}c).

It is instructive to consider the limit case of a two laser system in which laser 1 is frequency-stabilised and laser 2 is being frequency-tuned measuring a fixed distance $l$.  
If phase measurements taken at times $t_1$ and $t_2$ are analysed applying the PSS model, the PSS phase difference according to Eq.~(\ref{Eq_Synthetic_phases_for_PSS}) is equivalent to a 'synthetic phase' solely calculated from the optical phase of laser 2, as the constant phase information of laser 1 cancels out. Similarly, the resulting PSS wavelength is identical to the synthetic wavelength, which is derived exclusively from the two successive wavelengths $\lambda_2(t_1)$ and $\lambda_2(t_2)$. In practice, however, the inclusion of the constant laser 1 proves advantageous in terms of common-mode rejection, as mechanical vibrations – and the associated small displacements  in geometric lengths – cannot be completely suppressed in many relevant cases. Compared to operating at the optical scale, the NAR of the synthetic wavelengths significantly reduces sensitivity to unavoidable position changes. Furthermore,  the formalism requires knowledge of only the synthetic wavelength which is directly determined by the frequency offset between the two lasers rather than their absolute optical wavelengths. Consequently, the optical design of the source is considerably simplified, eliminating the need for absolute frequency stabilisation. The overall accuracy of the synthetic wavelength approach is ultimately determined by the quality of the offset locking between the two lasers. As fluctuations in the frequency difference directly affect the phase stability of the synthetic signal, tight control of the optical beat frequency is required. To achieve this, an optical phase-locked loop (OPLL) is used to actively track and suppress frequency fluctuations between the two optical signals, as explained in section \ref{Sec_Exp_Optical_Source} below.

\section{Experimental Setup} \label{Sec_Experimental}
To achieve field capability, the TeleYAG-II interferometer is divided into three functional units. The phase detection and laser stabilisation described in Sec.~\ref{Sec_Exp_Optical_Source}  form the foundation for the metrological principle of distance measurement via MWLI. Section~\ref{Sec_Exp_Interferometer_Head} details the field-capable optical interferometer design, which implements two-colour distance interferometry utilising spatially separated input beams delivered via photonic crystal fibres (PCFs). Section~\ref{Sec_Method_Signal_Demodulation} subsequently details the dedicated phase evaluation electronics and FPGA-based instrumentation designed to capture and extract phase signals from these heterodyne carriers. 

\subsection{The Optical Source: FPGA-Based Optical Phase-Locked Loop}\label{Sec_Exp_Optical_Source}
The TeleYAG-II absolute distance interferometer is based on two diode-pumped solid-state (DPSS) Nd:YAG ring oscillators (Innolight Prometheus 20NE), which provide fundamental-wave emission in the near-infrared (NIR) at 1064 nm with an optical power of 1 W and by second-harmonic generation (SHG) visible ligth (VIS) at \qty{532}{\nano\metre} with \qty{20}{\milli\watt}. The second harmonic generation (SHG) process ensures intrinsic optical phase stability relative to the fundamental wave. The low intrinsic noise and long coherence length of the Nd:YAG ring resonators enable observable interference over 100 km in fibre-optic delay lines \cite{Czarske1995,Sorin1990}, making these sources particularly well-suited for geodetic metrology. To ensure ease of use and robustness for field deployment, the optical assembly is operated without active vibration isolation or acoustic shielding.

Frequency stabilisation between the reference and follower lasers at \qty{1064}{\nano\metre}  is achieved via an OPLL. The principle is sketched in Fig.~\ref{Fig_concept_PSS}a.  Superimposed NIR light from the two Nd:YAG resonators is directed onto a fibre-coupled, high-speed photodetector (New Focus 1414, with a bandwidth of \qty{25}{\giga\hertz}) to generate a high-frequency optical beat signal. To enable subsequent processing of this signal in the digital control hardware, the approximately \qty{20}{\giga\hertz} beat signal is downconverted to an intermediate frequency (IF) of \qty{62.5}{\mega\hertz} using a double-balanced RF mixer driven by a low-noise SHF synthesiser (Rohde \& Schwarz SMA-100B, option SMAB-B710; single-sideband phase noise of  \qty{-111}{\dBc\per\Hz} at \qty{1}{\kilo\hertz} offset). The resulting IF signal is split by a power divider, enabling simultaneous real-time monitoring on an SHF spectrum analyser and digital processing within the feedback controller. Digital frequency detection and offset frequency control are performed on the same reconfigurable  field programmable gate array (FPGA) platform (NI-7935R, equipped with a Xilinx Kintex-7 XC7K410T). The \qty{62.5}{\mega\hertz} IF waveform is sampled at \qty{250}{\MegaSamples\per\second} by a 14-bit analogue-to-digital converter (ADC, type ADS62P49 on an NI-5782 transceiver module). A hardware integrated In-Phase/Quadrature (I/Q) demodulation algorithm extracts the instantaneous optical phase with a deterministic processing latency of approximately \qty{1}{\micro\second}. The derived frequency error signal is then processed by a 256-tap finite impulse response (FIR) filter, whose passband can be configured down to \qty{125}{\kilo\hertz} in order to detect and suppress disturbances while ensuring the stability of the control loop. To achieve both fast frequency tracking and a wide dynamic range, the filtered error signal is fed into a cascaded PID controller topology comprising two control paths. High-frequency phase fluctuations are dynamically suppressed via the fast control path. Here, a 16-bit, high-speed digital-to-analogue converter (DAC, type DAC5682Z with a \qty{77}{\dB} spurious-free dynamic range (SFDR) and a\qty{163}{\nano\second} latency) directly modulates the piezoelectric transducer (PZT), which is attached to the follower laser crystal. Meanwhile, slow thermal frequency drifts are compensated for via a parallel 20-bit precision DAC (AD5791, \qty{100}{\dB} SFDR and \qty{1}{\micro\second} latency), which controls the temperature of the follower laser crystal and thus prevents the PZT's measurement range controller from saturating. An observation time of \qty{1}{\second} is well suited for evaluating frequency stability with respect to range measurement results. The phase-locked \qty{20.0625}{\giga\hertz} yields an average linewidth of approximately \qty{150}{\hertz}. With a static offset frequency of \qty{20}{\giga\hertz}, the OPLL sets fixed primary SWLs of \qty{14.94}{\milli\metre} (\qty{1064}{\nano\metre}) and \qty{7.47}{\milli\metre} (\qty{532}{\nano\metre}). This remaining frequency uncertainty of \qty{150}{\hertz} corresponds to a relative stability of $7.5 \times 10^{-9}$.

Bridging the scale from millimetre synthetic wavelengths to geodetic ranges is achieved by continuously tuning the follower laser offset frequency by using the analogue ramp-sweep function (option: SMAB-28) of the SHF synthesiser. To automate this process, we control the synthesiser via Standard Commands for Programmable Instruments (SCPI). This enables the execution of highly reproducible frequency sweep sequences during operation, allowing subsequent offline data processing to be handled by a standardized script without requiring manual adjustments for individual runs. In this mode, stable phase values can be determined over several seconds after the target frequencies have been reached. To generate heterodyne signals for laser channel separation, the TeleYAG-II features a set of six fibre-coupled acousto-optic modulators: three for VIS (Brimrose BRI-TEF-200-532-2FP) and three for NIR (Gooch \& Housego SFO1915-T-M200-0.1C2G-3-F2P). These AOMs operate at nominal carrier frequencies around \qty{200}{\mega\hertz} and, when superimposed, produce slightly different frequency offsets between \qty{1.95}{\mega\hertz} and \qty{3.72}{\mega\hertz}. The NIR and VIS beams are combined using a dichroic plate before being coupled into two photonic crystal fibers (PCFs), with each of these fibers delivering optical powers of \qty{10}{\milli\watt} (NIR) and \qty{0.3}{\milli\watt} (VIS) directly to the interferometer head.

\subsection{Interferometer Head} \label{Sec_Exp_Interferometer_Head}
\begin{figure}
 \centering
	\includegraphics[width =1\columnwidth]{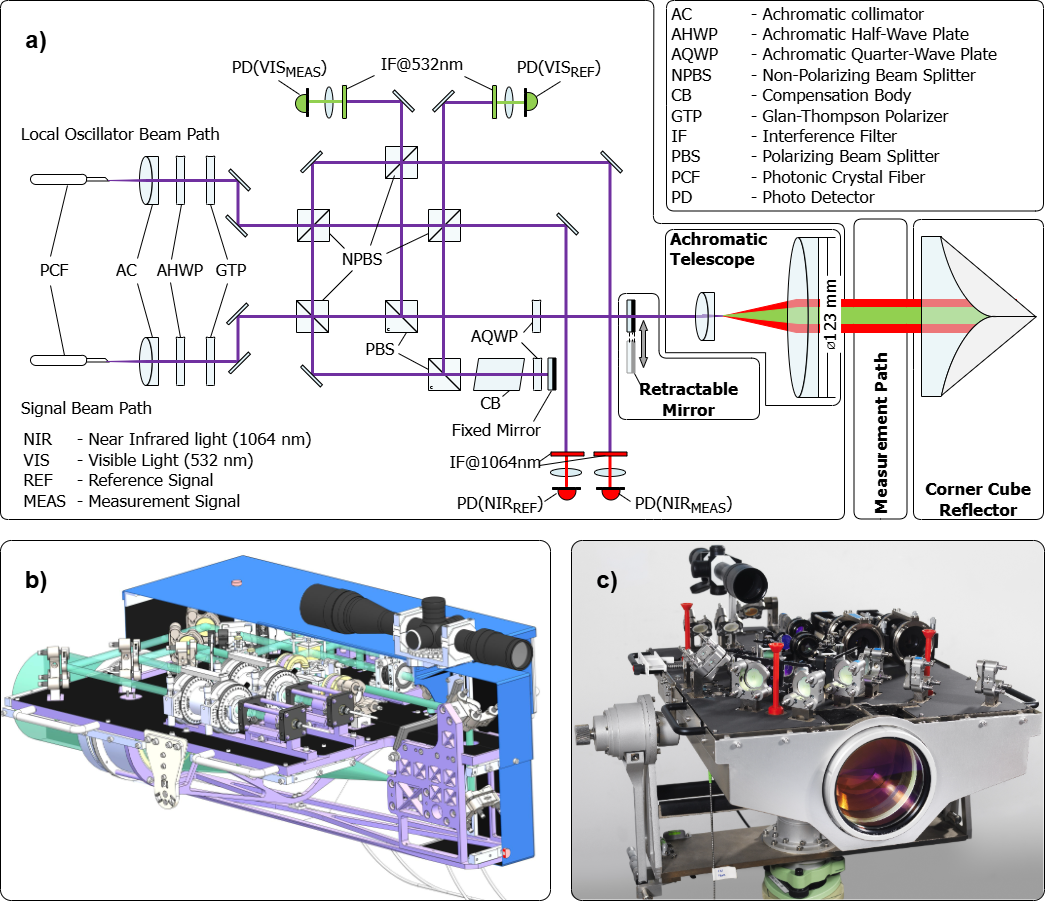} 
 \caption{a) Functional scheme of the heterodyne two-colour absolute distance interferometer TeleYAG-II. b) CAD side view through the interferometer. c) Front view of the assembled interferometer.}
\label{Fig_setup_TeleYAG_II}
\end{figure}
The purpose of the TeleYAG-II interferometer is to serve as primary standard for baseline calibrations of ranges up to \qty{5}{\kilo\metre}. Unlike previous laboratory demonstration instruments like it predecessor TeleYAG \cite{MeinersHagen2015, MeinersHagen2017}, the interferometer shall be able to work under typical outdoor conditions continuously without a constant need for realignment by the operators on baselines all over Europe. This defines a number of practical requirements for the realisation. For once, the system must endure a temperature range of \qty{0}{\celsius} to \qty{30}{\celsius}. Extended integration times are necessary to reduce the impact of turbulence. Therefore, the design must maintain structural, in particular thermal stability even over several tens of minutes on micrometre level. Low-strain mounting of the optical components of the interferometer with very uniform force vectors to ensure long-term stability is also important for mechanical robustness against shock or vibrations during transport and setup. Further to this,  the system centre of gravity shall coincide with the origin of the three mechanical axes because torques due to load mass at the geodetic mounting point must be avoided. In addition, the emitted laser beam must be aligned to a retro-reflector with a circular aperture of only \qty{127}{\milli\metre}. This translates for a design length of up to five kilometres to an alignment precision of less than 0.5 arcsecond in both axes. Moreover, the system should have a well-defined and stable reference point, ideally insensitive to typical optical alignments. Last but not least, two persons need to be able to mount the instrument on geodetic pillars.  For a better handling during transport and installation, the system is divided into two units: the interferometer head, which accommodates all parts of the optical set-up, and the base with the alignment setup. The latter is compatible to standard geodetic tribrach mounts. Repeatability of the mounting is ensured by a  self-centring mechanism between the two units. One side of the interface has three pairs of dowel pins acting as V-prisms, while the other side consists out of three spheres. They are aligned on the same reference diameter in a \qty{120}{\degree} angle. Mounting both sides together with a central screw leads to a very well defined and kinematically determined bearing with high reproducibility in the alignment of the optical and the mechanical axes \cite{Pilarski2021}. 

The basic optical design of the TeleYAG-II interferometer depicted in Fig.~\ref{Fig_setup_TeleYAG_II}a is that of a heterodyne interferometer setup. The optical components are mounted on a stiff closed-frame made of an Invar alloy (see Fig.~\ref{Fig_setup_TeleYAG_II}b and \ref{Fig_setup_TeleYAG_II}c) with low coefficient of thermal expansion ($\alpha \approx \qty{1.2e-6}{K^{-1}}$). All optical mounts are selected for robustness. The angles between optical components and the holding fixtures are optimised to avoid parasitic etalon effects in the interferometer. In its targeted final stage, the TeleYAG-II interferometer shall be able to compensate the index of refraction by the two-colour method \cite{MeinersHagen2017}.  VIS and NIR light beams  are concentrically emitted by the polarisation-maintaining photonic crystal fibre (PM-PCF, AlphaNOV LMA-PM-5 FC/APC - FC/PC). The beam waists of the two harmonics, however, differ by approximately a factor of two \cite{Liu2020}. To avoid the tedious manual beam superposition alignment on the head, VIS and NIR beams are to remain superposed all the time on the TeleYAG-II interferometer head. All optical elements are anti-reflection coated for the VIS and for the NIR wavelengths. Local oscillator (LO) and signal beams are fed to the interferometer head by the two PM-PCFs from the optical source (see Sec. \ref{Sec_Exp_Optical_Source}). They are collimated by two specifically designed two-lens achromats (Bernhard Halle Nachfolger (BHN) 2019.0061.0004). The polarisations of LO and signal beams are subsequently rotated by half wave plates (BHN 2017.0167.0003) and  purified by Glan-Taylor-polarisers (Thorlabs GT15-YAG-SP). They are then both distributed by two non-polarising beam splitter plates into the respective reference and measurement paths of the interferometer.  Incoming and returning beams of reference and measurement paths are routed by polarisation-based optical switches consisting of a polarising beam splitter and a quarter wave plate (BHN 2019.0057.003 and BHN 2017.0150.003). A glass body is incorporated into the reference path to balance the telescope glass paths. We have designed special mounts for beam splitters and wave plates that ensure uniform and low-stress restraint of the optical components to conserve beam polarisation. As Gaussian beam, the collimated beam varies with propagation length. To reduce this impact, the beam is enlarged by a custom achromatic telescope of a two-lens ocular (BHN 2019.0061.0001) and a three lens objective (BHN 2019.0061.0002). The beam waist of the VIS measurement beam is expanded to approximately \qty{15.006}{\milli\metre} after the objective. The beam diameters are chosen that for ranges of up to \qty{1}{\kilo\metre} incoming and returning beam diameters match  \cite{Sauthoff2023}.  The optics spacing between ocular and objective lens is approximately \qty{560}{\milli\metre}. To keep the system compact, the telescope is mounted on a separate level below the interferometer optics (Fig.~{\ref{Fig_setup_TeleYAG_II}b). The measurement beam is guided by a two mirror telescope into the lower level of the interferometer mount. As retro-reflector, a five inch silver coated retro reflector with a beam deviation smaller than \qty{0.5}{\arcsecond} is used (PLX-USHM-50-05E). Its partial mirrors are mechanically wrung as glued fixings have turned out to quickly degrade in the humid environment outdoors.  After being reflected, both signal beams are superposed with the respective LO beams and interfere in non-polarising beam splitters (Laser components). Fifty percent of the signal passes a 532 nm bandwidth filter and is coupled into a \qty{200}{\micro\metre} core multi-mode fibre (MMF, type Diamond FG200LEA FC/PC-FC/APC). The instrument enclosure is thermally as well as mechanically isolated from the optical set-up to eliminate the influence of the surroundings during the measurement (see Fig.~\ref{Fig_setup_TeleYAG_II}b). Beam alignment and targeting over hundreds or even thousands of metres is realised with a three-step alignment procedure. In a first step, the instrument is coarsely aligned by hand and clamped. The second stage consists of a small linear table which moves a micrometer gauge. After fixation of the second stage, a fine-tuning gauge connected to a lever finally enable the instrument to be aligned very precisely with the target reflector. 

\subsection{Signal Demodulation and Phase Extraction} \label{Sec_Method_Signal_Demodulation}
As discussed in  Sec.~\ref{Sec_Method}, the absolute distance measurement realised by the TeleYAG-II system is based on the measurement of the optical phases $\phi_1$ and $\phi_2$. The respective beams are spatially perfectly superposed (see Sec. \ref{Sec_Exp_Interferometer_Head}) and their VIS wavelengths with \qty{532.24}{\nano\metre} and \qty{532.20}{\nano\metre} nearly identical. Such signals can be separated with a minimal set of photo detectors, i.e.  one set for  measurement and reference beams in the VIS and NIR, respectively \cite{Yang2015, Koechert2012}.  Fibre coupling enhances the robustness of the overall system and, thus, improves its field capability by protecting delicate optical alignments from environmental disturbances. After passing through the interferometer optics and interference, reference and measurement signals are guided by  MMFs to two fibre-coupled photo receivers per colour channel which are not mounted on the head (FEMTO Messtechnik HCA-S-200M series, -SI-FC and -IN-FC version für VIS and NIR, respectively). The heterodyne beat signals are digitised by a multichannel phase meter (NI-5762 ADC module, \qty{75.5}{\dB} signal-to-noise ratio (SNR)).  The phase meter introduces a noise corresponding to \qty{5}{\pico\metre} or better. This corresponds to an intrinsic phase resolution of $\delta \phi^{\rm int} =  \qty{0.13}{\milli \radian}$.  For further processing, the raw digitised data are stored as binary-encoded files, converted to ASCII format, and then post-processed offline using Python scripts to demodulate the heterodyne signals and perform further (pseudo-)phase unfolding across the entire measurement range.

\section{Experimental Results}\label{Sec_Results}
In this section, we present experimental data from a dedicated measurement campaign for the system characterisation. In Sec.~\ref{Sec_Results_RawData}, we discuss the noise properties of the raw, unaveraged interferometric data. In Sec.~\ref{Sec_Results_Sweeping} we apply the TSWL and PSS approach to determine the absolute coarse length. These are used to unwrap the final high-resolution observation results in Sec.~\ref{Sec_Results_HighResolution}.
\subsection{Raw Data}\label{Sec_Results_RawData}

The interferometric measurements discussed in this paper were performed during a measurement campaign in autumn 2022 at the novel \qty{250}{\metre} baseline at Warsaw, Poland \cite{Pollinger2023}. The interferometer was positioned on pillar 03 of the baseline. Measurements were performed to pillar 00 with a nominal baseline $l_1$ of approx. \qty{28}{\metre} and to pillar 07 with a nominal baseline $l_2$ of approx. \qty{224}{\metre}. The data were acquired within twenty-four hours. The shorter baseline was measured in the afternoon in partial sunlight. The data for the longer baseline were mainly collected at night and during the subsequent morning, when there was partial cloud cover and rising winds. The ambient conditions were monitored by a mobile weather station. Five temperature sensors were distributed along the baseline at beam height, and two sensors above and below this line to monitor gradients. Humidity and ambient pressure were monitored with one dedicated sensor each. 

The environmental data are recorded every ten seconds. The distance data are captured at a frequency of \qty{3.8125}{\kilo\hertz}. For an SWL measurement, 2.3 million data points are taken for the shorter, 2.3 to 2.7 million for the longer baseline. The phase meter measures the quadrature components ($\sin\phi_i, \cos\phi_i $) of the two visible laser lines \cite{Pollinger2026}. The data are normalised and a Heydemann correction \cite{Heydemann1981, Koening2015} is applied on ten subsets of the datasets to correct for possible remnant non-linearities. With the help of the ${\rm arctan2}$ function, the phase values modulo $2\pi$ are subsequently recovered. We calculate the synthetic phases $\phi^{\rm swl}_j$ from the corrected phase data according to Eq.~(\ref{Eq_SynPhase}), and the optical lengths derived according to  Eq.~(\ref{Eq_LengthFrom2Colour}). For refractivity compensation, temperature, pressure and humidity sensor data are logged every ten seconds and linearly interpolated to match the data acquisition rate. Using the CiddorPy module \cite{Pollinger2020}, the phase refractive indices for the two optical wavelengths are derived from the environmental data, and consequently the two-colour refractive index $n^{\rm 2WL}$ calculated according to Eq.~(\ref{Eq_Two_colour_index_of_refraction}). The measured optical paths are then corrected for the index of refraction. Figure \ref{Fig_RawData}a shows exemplary SWL raw data as deviation from their mean value for three measurements, one for the shorter distance $l_1$ taken during the day (i), and for the longer distance $l_2$ a day (ii) and a night (ii) measurement \cite{Pollinger2026}. The data for the shorter distance $l_1$ show a low level of background noise of approximately 0.2 mm. Isolated peaks can be explained, for example, by insects in the beam path. The quality of the data for the longer distance $l_2$ varies considerably between the two measurements shown. The daytime measurement (ii) shows strong, persistent noise components of 1.5 to 2.0 mm. Such an increase compared with the shorter distance is, of course, to be expected due to the longer air path. The sensor density – and thus the quality of the compensation – decreases for the longer distance, whilst at the same time the influence of turbulence increases with distance.  Somewhat more surprising, however, is the quality of the measurement data for the long distance recorded during the night (iii). The noise level is significantly lower than in both daytime measurements. The measurement conditions were evidently extremely favourable, with the atmosphere very stable. It is only towards the end of the measurement series that greater variation occurs. At this point, light rain began to fall during the measurement.

\begin{figure}
 \centering
	\includegraphics[width =1\columnwidth]{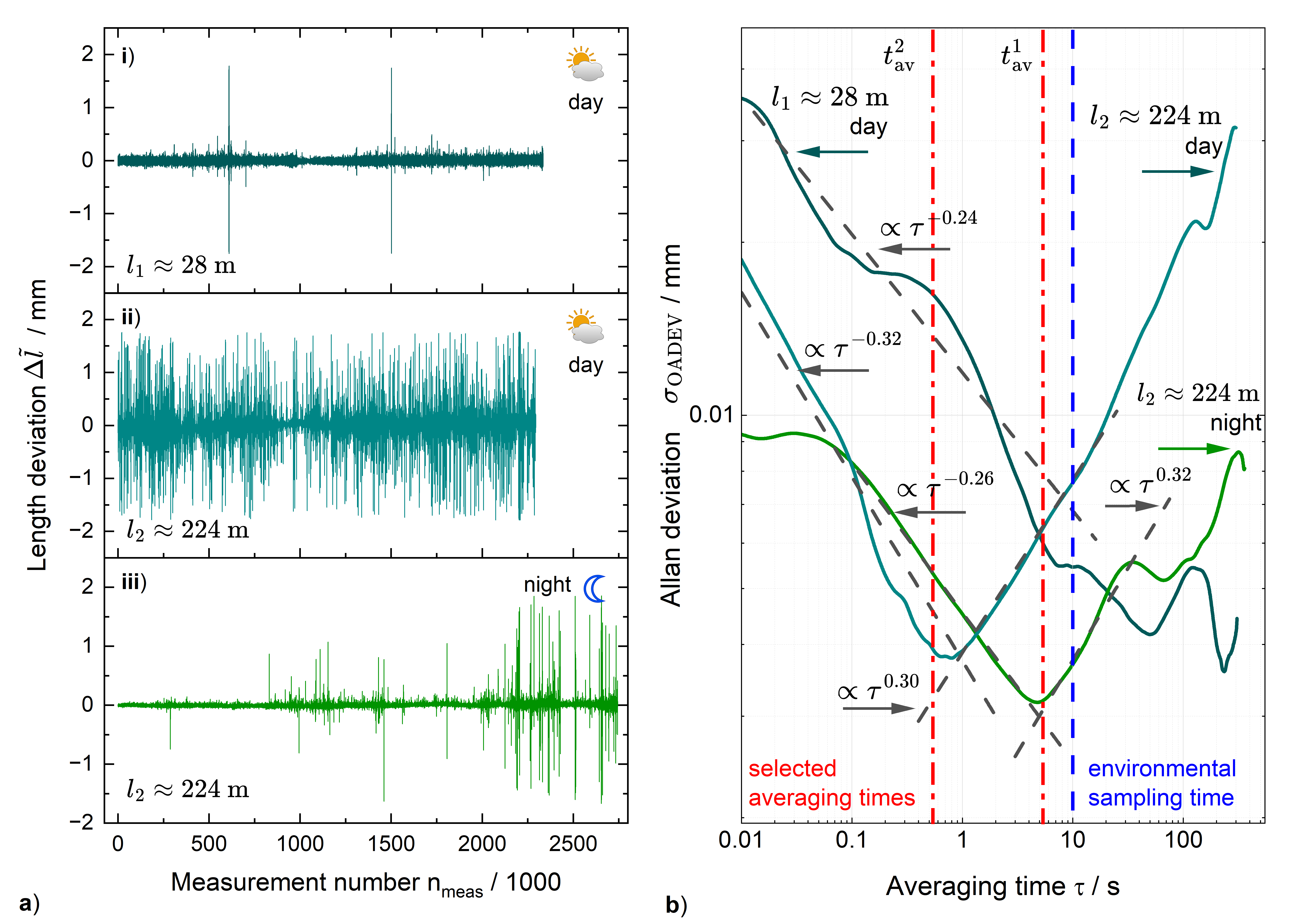} 
 \caption{Raw SWL observation data for $l_1 \approx \qty{28}{\metre}$ and $l_2 \approx \qty{224}{\metre}$ taken at day and nighttime. a) Deviation of the raw data points from the respective mean values. b) Overlapping Allan deviations.}
\label{Fig_RawData}
\end{figure}

The measurement is affected by various sources of noise, the typical time scales of which vary. Vibrations may occur in the structure supported by the pillars. Furthermore, mechanical drifts in the structure cannot be entirely ruled out during measurements lasting several minutes. However, given the structural stability of the TeleYAG-II system, the influence of the atmosphere is considered to be significantly greater. The temporal resolution of the weather station data is limited not only by the sampling interval but also by the latency of the sensors. Furthermore, the spatial sampling of the environmental parameters is not perfect. This leads to high- and low-frequency noise components. Atmospheric turbulence causes lateral displacement of the measurement beam. As the position of the reference beam in the interferometer head remains constant, the relative position of the reference wavefront and the measurement wavefront shifts relative to one another during interference, leading to phase shifts. Furthermore, the efficiency of the fibre coupling into the MMF connecting the interferometer head to the detectors varies. The resulting amplitude modulation can also lead to phase modulation in the signal. Besides, the standard deviation of the OPLL frequency also leads to high-frequency noise in the distance data. The overlapping Allan deviation $\sigma_{\rm OADV}$ allows identification and separation of different drift sources and suitable integration times \cite{Howe1981, Shmagun2022}.  It is depicted for all three datasets in Fig.~\ref{Fig_RawData}b \cite{Pollinger2026}. All three curves decline at a similar rate (-0.24 to -0.32). For the short distance, this trend is essentially maintained throughout the entire observation period. For the longer distance, the two-sample variance rises again at averaging times exceeding 0.7 s for the daytime measurement and approximately 5 s for the nighttime measurement. The rate of increase is comparable in both cases at approximately 0.31. Recent work by Rustom et al. \cite{Rustom2026} provides a qualitative understanding of the major trends. For shorter integration times, the influence of turbulence on phase measurement predominates, whilst the increase in the Allan variance at longer integration times can be explained by uncompensated refractive index compensation and mechanical drifts. Clearly, the latter factor carries greater weight for the longer distance within the observed time window than for the shorter one, which, relatively speaking, is obviously monitored more effectively given the same number of sensors. During the day, greater inhomogeneities in the refractive index, as well as a greater thermal influence on the components, are to be expected. It is therefore not surprising that, in the daytime measurement, the increase in the Allan variance begins at shorter averaging times than in the nighttime measurement. Nevertheless, it is somewhat surprising that the two-sample variance for the short distance is slightly greater for shorter integration times than in both measurements of the longer distance. A greater wind load, or perhaps slightly poorer optical alignment, may have played a role here. 

From a practical metrological point of view, the Allan variance enables us to identify the ‘sweet spot’ for an optimal integration time. The aim is to suppress stochastic noise whilst ensuring that systematic drifts are not masked. Another factor influencing the choice of an optimal integration time is also indicated in Fig.~\ref{Fig_RawData}b. The 10-second-sampling interval for ambient measurements  indicated by the blue dashed line in Fig.~\ref{Fig_RawData}b determines  the time scale for correcting refractive index drift and sets a Nyquist limit of \qty{50}{\milli\hertz} to resolve drift behaviour. To achieve maximum noise reduction without over-smoothing the physical target dynamics or losing the slower trends, we choose to average over $N_{\rm av}^1= 20,480$ samples for all high resolution observation data. This window corresponds to an integration time  $t_{\rm av}^{1}$ of approx. \qty{5.4}{\second} as indicated by the red dashed line in Fig.~\ref{Fig_RawData}b. For the shorter observation times during the TSWL measurements, which were performed mostly during daytime, we choose a shorter integration time $t_{\rm av}^{2}$ of approx. \qty{0.54}{\second}. In the following We use a bin-average approach for calculation efficiency, which replaces dataset bins of $N_{\rm av}^{1}$ points by their median value. This averaging acts as a low-pass filter with a 3-dB bandwidth of approximately $\qty{84.4}{\milli\hertz}$ and attenuates the spectral noise above the cut-off frequency, as well as the measurement noise. 

\subsection{Tunable Fixed Synthetic Wavelength and Pseudo-Supersynthetic Wavelength Interferometry} \label{Sec_Results_Sweeping}
\begin{figure}
 \centering
	\includegraphics[width =1\columnwidth]{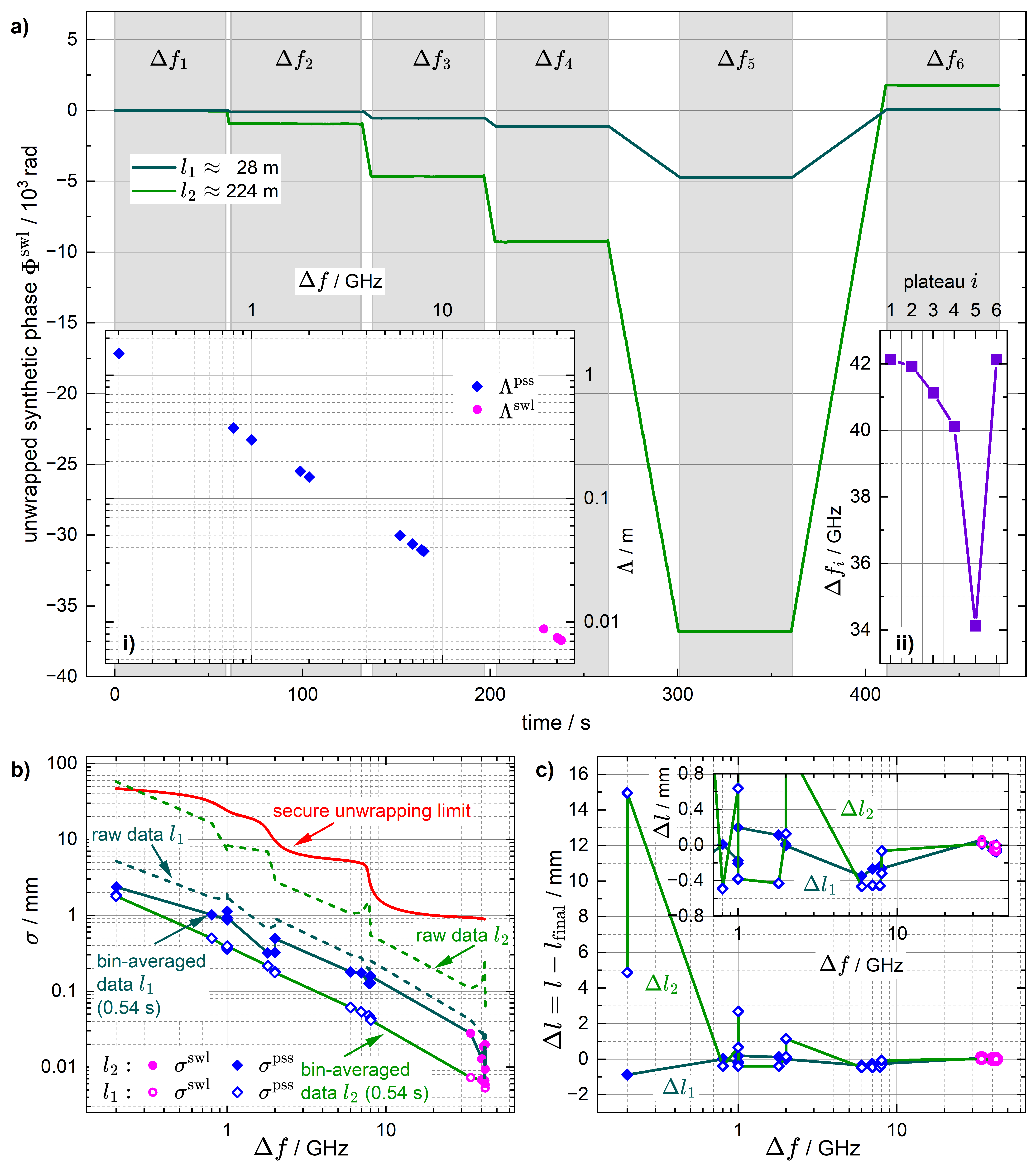} 
 \caption{Absolute distance measurement by TSWL and PSS interferometry. a) Unwrapped unaveraged raw synthetic phase data of the TSWL experiment. i) Resulting pseudo-synthetic and synthetic wavelengths. ii) OPLL frequency steps for the TSWL measurement. b) Standard deviations of the PSS and SWL length measurement results compared to their maximum permissible value for a secure unwrapping of the subsequent measurement result. c) Iterative unwrapping of the length measurement. It is worth noting that cycle slips, such as those present in the green plot shown in (a), can happen due to frequency jumps during the tuning process. However, they are not relevant for the length analysis because only the phase information from the stable plateaus is considered.}
\label{Fig_Sweep}
\end{figure}

Figure~\ref{Fig_Sweep} shows  exemplary measurement data obtained applying TSWL and PSS  interferometry \cite{Pollinger2026}. The frequency difference $\Delta f$ between the two emitted visible Nd:YAG laser lines is varied in six steps between \qty{42.2}{} and \qty{34.1}{\giga \hertz} as depicted in the inset (ii) of Fig.~\ref{Fig_Sweep}a during a total observation time of approximately \qty{470}{\second}.  The optical phases $\phi_1(t)$ and $\phi_2(t)$ are continuously recorded approximately every \qty{0.31}{\milli\second}. Once a targeted frequency difference $\Delta f_i$ is reached, the laser frequency is kept fixed for \qty{60}{\second}. These plateaus $\Delta f_1, \Delta f_2, ...$ to $\Delta f_6$ are marked in grey in the background of the main graph of Fig.~\ref{Fig_Sweep}a. The data presented there depict the unaveraged unwrapped synthetic phase $\Delta\Phi^{\rm swl}(t)$ according to Eq.~(\ref{Eq_SynPhase}) of two measurements at $l_1 \approx \qty{28}{\metre}$ and $l_2 \approx \qty{224}{m}$, both taken at daytime.  Combining Eq.~(\ref{Eq_Length_From_Synthetic_WL}) and (\ref{Eq_Def_Synthetic_Wavelength_by_frequency}), the absolute phase change  $\Delta\Phi^{\rm swl}(t)$ is expected to be proportional to the product of frequency difference $\Delta f$ and the optical path length $l$. The data in Fig.~\ref{Fig_Sweep}a shows a qualitative linear dependence of the phase change from the frequency difference $\Delta f_i$. Furthermore, the observed change for the \qty{224}{\metre} is also roughly seven times larger than for the \qty{28}{\metre} distance. But the data also shows that the experiment is not perfect, and a naive exploitation of Eq.~(\ref{Eq_Length_From_Synthetic_WL}) and (\ref{Eq_Def_Synthetic_Wavelength_by_frequency}) based on the unwrapped synthetic phase  $\Delta \Phi^{\rm swl}$ is bound to fail. The frequency variation cycle finishes with a repetition of the measurement at $\Delta f_6 = \qty{42.2}{\giga\hertz} = \Delta f_1$ as depicted in inset ii) of Fig.~\ref{Fig_Sweep}a. In theory, the synthetic phase $\Delta \Phi^{\rm swl}_6$ is expected to return back to the initial value $\Delta \Phi^{\rm swl}_0$ due the identical frequency difference and distance. But this is not the case for both distances. The $l_2$-data clearly overshoot the zero line. The final plateau $i=6$ deviates by \qty{1797.4}{\radian} from the initial value \qty{0.4}{\radian}. This difference in absolute phase corresponds to a nominal change in distance by \qty{1.02}{\metre}.  It can obviously be excluded that the baseline is changing by this amount during the observation. The reason for the imperfect 'closed circle' is rather the fact that the unwrapping process fails due to imperfect sampling during the linear frequency variation. Over the complete experiment, multiple cycle slips occur. Generally speaking, they indicate that frequency changes are occurring too rapidly whilst the sampling rate is too low. A  direct interpretation of the absolute unwrapped synthetic phase shift due to the frequency variation is hence doomed to fail if neither the frequency change is substantially decreased nor the data acquisition fundamentally increased. While the latter is limited by technical and economical considerations, the former increases the impact of mechanical instabilities. Moreover,  the situation is further complicated in this specific case by the fact that the SHF generator used does not allow for continuous frequency changes, particularly across large frequency steps. During the sweep, short plateaus must be maintained repeatedly to allow self-referencing, which lead to non-linear frequency responses in the OPLL control system. In summary, a direct interpretation of the absolute phase difference $\Delta \Phi(\Delta f)$ in dependence of the frequency change is not robust enough for a reliable absolute distance measurement under these conditions.

The TSWL and PSS concept introduced in Sec.~\ref{Sec_Method}, however, allows a more robust absolute distance measurement using the identical information.  As prerequisite, a series of pseudo-synthetic and synthetic wavelengths $\Lambda^{\rm pss}_i$ and $\Lambda^{\rm swl}_j$ are required for which the uncertainties of the corresponding length measurement fulfil the secure unwrapping limit given by Eq.~(\ref{Eq_Condition_PSS_Syn_Series_uncertainty_lambda_fourth}). Inset i) of Fig.~\ref{Fig_Sweep}a shows the synthetic and pseudo-supersynthetic wavelengths that can be constructed from the different stable difference frequency plateaus in the measurement. The classic SWLs $\Lambda^{\rm swl}_i$ associated with the six plateaus can be calculated by Eq.~(\ref{Eq_Def_Synthetic_Wavelength_by_frequency}). They are represented by pink full circles. The shortest SWL $\Lambda^{\rm swl}_{\rm min}$ is given by \qty{7.12}{\milli\metre} while the longest increases only to  \qty{8.79}{\milli\metre}, only slightly extending the nominal NAR. By the construction of PSS wavelengths $\Lambda^{\rm pss}_j$ using Eq.~(\ref{Eq_Def_PSS_Wavelength_From_Delta_fs}), however, the NAR can be considerable enlarged. The PSS wavelengths $\Lambda^{\rm pss}_j$ obtainable in the chosen frequency combinations vary between \qty{3.75}{\centi\metre} and \qty{1.499}{\metre}.  They are indicated by solid blue diamonds in inset i) of Fig~\ref{Fig_Sweep}a. The secure unwrapping criterion (\ref{Eq_Condition_PSS_Syn_Series_uncertainty_lambda_fourth}) demands that the uncertainty of a distance measurement of a given element in the measurement series must not be larger than a fourth of the synthetic (or pseudo-synthetic) wavelength of the subsequent interferometric measurement in the chain. As discussed in Sec.~\ref{Sec_MU}, there are a number of systematic uncertainty contributions to the SWL measurement that are independent from the frequency difference or dependent on the environment only. If these are subtracted from the maximum permissible value given in Eq.~(\ref{Eq_Condition_PSS_Syn_Series_uncertainty_lambda_fourth}), the maximum permissible standard deviation for a secure unwrap remains. This value for each SWL interferometry measurement is indicated by the the solid red line in Fig.~\ref{Fig_Sweep}b.  The green dashed lines indicate the sample standard deviation of the raw length measurements for the respective PSS and SWL interferometry measurements. The SWL standard deviations $\sigma^{\rm swl}_i (l)$ are derived from the observed standard deviations of the synthetic phase $\tilde{\sigma}^{\rm swl}_i(\phi^{\rm swl}_i)$ according to 
\begin{equation}
\sigma^{\rm swl}_i (l) = \frac{\tilde{\sigma}^{\rm swl}_i(\phi^{\rm swl}_i)}{4 \pi n^{\rm 2WL}_i} \Lambda^{\rm swl}_i. \label{Eq_Standard_Dev_Exp_Syn_Data}
\end{equation}
The PSS result, however, is not directly observed as two temporally subsequent measurement results are combined by using their median values $\left< \phi^{\rm swl}_i \right>$ and $\left< \phi^{\rm swl}_j \right>$  in Eq.~(\ref{Eq_Def_PSS_Phase}) and  (\ref{Eq_Length_From_PSS_Measurement}). As a consequence, the standard deviation of the PSS length measurement results $\sigma^{\rm pss}_i (l)$ can be calculated using
\begin{equation}
\sigma^{\rm pss}_{i,j} (l) = \frac{\sqrt{\tilde{\sigma}^{\rm swl}_i(\phi^{\rm swl}_i)^2+\tilde{\sigma}^{\rm swl}_j(\phi_j)^2}}{4 \pi n^{\rm pss}_{i,j}} \Lambda^{\rm pss}_{i,j}. \label{Eq_Standard_Dev_Exp_PSS_Data}
\end{equation}
Even for the standard deviation of the raw data, all but one experimental values remain below the maximum permissible value.  The solid lines in Fig.~\ref{Fig_Sweep}b indicate the standard deviation of data bin-averaged for an observation time of $t_{\rm av}^2 = \qty{0.54}{\second}$, chosen for the optimum integration time for the daytime $l_2$ distance measurement (see Fig.~\ref{Fig_RawData}b). For both datasets, the standard deviations substantially decrease. It might be surprising that the effect is considerably larger for the longer distance measurement. In this case, air turbulence has a stronger influence than for the 'shorter' distance, implying a higher standard deviation of the individual observation.  This effect, however, obviously induces statistical noise that averages out on the chosen time scale of a second. It should be noted that the standard deviations of the bin-averaged data are comparable for both cases.  The fact, however,  that the standard uncertainty of the longer distance presents itself even slightly smaller than in case of the shorter baseline should not be over-interpreted. Such minor differences can be explained by a slightly less optimum parameter choice for the OPLL or slightly varying mechanical stability due to wind or sun exposure during the observation, for example. But a secure unwrap can obviously be ensured by distance observations over a second in this case.  Figure~\ref{Fig_Sweep}a finally demonstrates the convergence of the length measurements along the synthetic wavelengths series. For the given frequency differences $\Delta f_i$, the longest constructable PSS wavelength is given by \qty{1.499}{\metre}. Hence a pre-value with an uncertainty of \qty{375}{\milli\metre} is needed for a correct unwrap of this largest scale. If such a value was not available the variable synthetic wavelength measurement could be easily extended by plateaus with frequency shifts $\Delta f$ smaller than \qty{200}{\mega\hertz}. For the stable baseline investigated here, such additional measurements are not necessary. Within the series of consecutive synthetic interferometry measurements, the derived length oscillates around the final value, the deviation amplitude decreasing with the synthetic wavelength $\Lambda^{\rm pss}_j$ and $\Lambda^{\rm swl}_i$. The procedure leads to solutions stable to \qty{200}{\micro\meter}-level for the optical path length which includes the geometric inter-pillar distance as well as the dead path of the interferometer. 

\subsection{High-Resolution Distance Measurement} \label{Sec_Results_HighResolution}
\begin{figure}
 \centering
	\includegraphics[width =1\columnwidth]{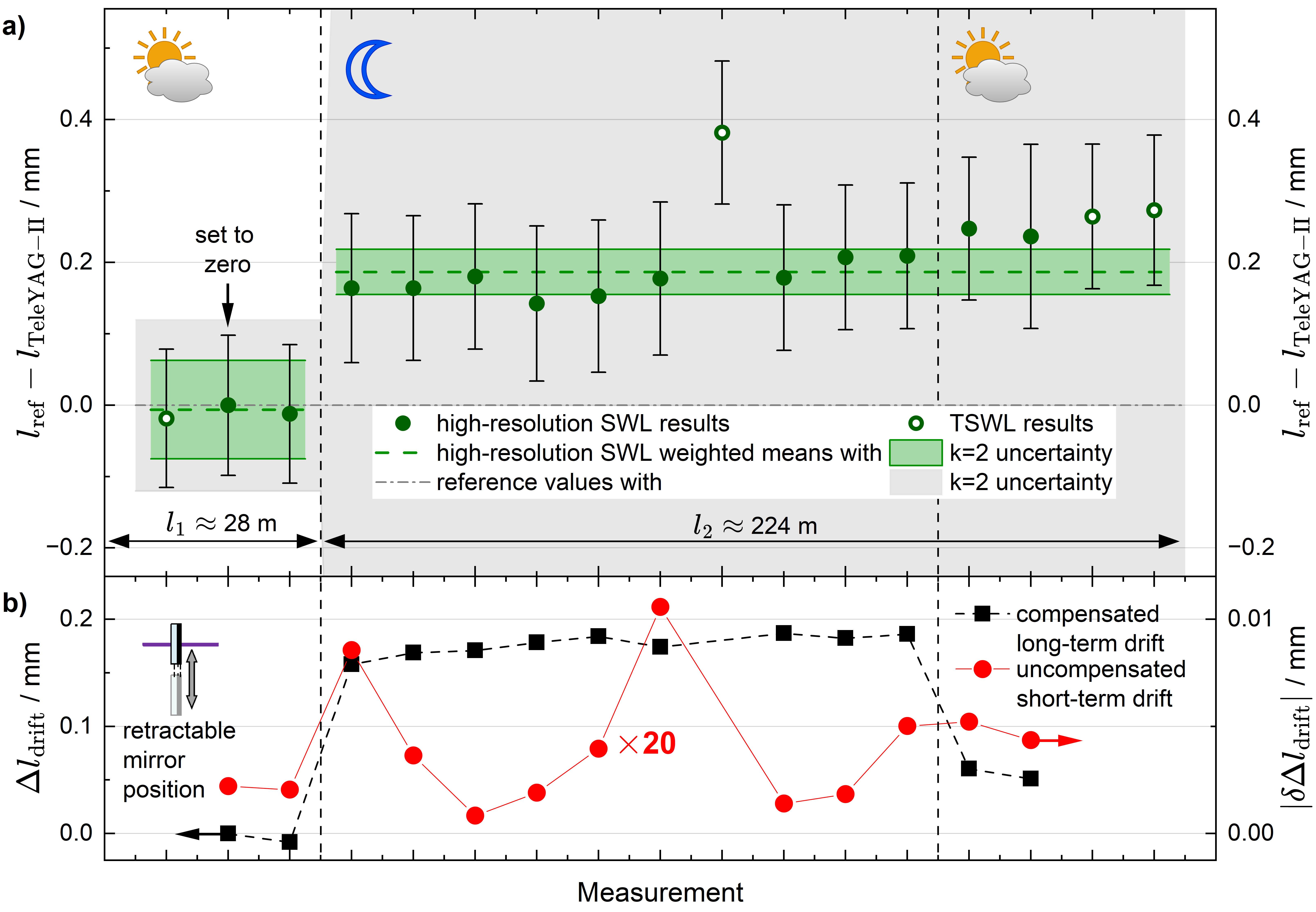} 
 \caption{a) Compilation of all TSWL and high-resolution SWL results and the derived mean values as deviations from the reference value. b) Compensated long-term drift $\Delta l_{\rm drift}$ and uncompensated short-term drift $|\delta \Delta l_{\rm drift} |$  as monitored by the internal reference mirror.}
\label{Fig_FinalObsResult}
\end{figure}

As mentioned, the fundamental measurement principle is the determination of the length $l$ based on Eq.~(\ref{Eq_LengthFrom2Colour}). The lowest uncertainty can be expected for the shortest synthetic wavelength $\Lambda^{\rm swl}_{\rm min}$ in the system.   The target of the TSWL experiment step discussed in Sec.~\ref{Sec_Results_Sweeping} is the determination of a pre-value $l^{\rm coarse}$ with an uncertainty $u(l^{\rm coarse})$ sufficiently small to fulfill the secure unwrapping criterion (\ref{Eq_Condition_Series_uncertainty_lambda_fourth}) for the shortest synthetic wavelength $\Lambda^{\rm swl}_{\rm min}$ in the system. Then the integer order $N_k$ in Eq.~(\ref{Eq_Unwrap_Order_Number}) can be determined. The coarse length value is now refined by a set of dedicated observation SWL measurements at a fixed frequency difference of $\Delta f_{\rm max} = \qty{42.125}{\giga\hertz}$, corresponding to a synthetic wavelength $\Lambda^{\rm swl}_{\rm min}$ of approximately  \qty{7.12}{\milli\metre}.  While the observation on the individual plateaus in the sweep measurement depicted in Fig.~\ref{Fig_Sweep}a takes about eight minutes in total and one minute per frequency difference $\Delta f_i$, the observation for the high-resolution measurements takes between eleven and twenty-one minutes. Immediately before and after the respective observation time the system's reference position is determined by a measurement of the retractable mirror position for one minute each. 

Figure~\ref{Fig_FinalObsResult}a summarises all measurement results for both baselines $l_1$ and $l_2$ in one figure \cite{Pollinger2026}. The green circle symbols represent the median values of the \qty{5.4}{\second}-bin-averaged data. The full circles indicate results obtained by the high-resolution SWL measurement. The final results of the coarse measurements obtained by TSWL interferometry are also included for completeness.  Two high-resolution SWL and one coarse TSWL measurement cycles performed at daylight are available for the shorter distance $l_1$, while there are nine  SWL and one TSWL nighttime and two SWL and two TSWL daytime measurements. The error bars represent the expanded measurement uncertainty $U(l)$ with a coverage factor $k = 2$ for the final result, covering both experimental standard deviation and systematic uncertainty contributions. The measurement uncertainty is discussed in detail in the following section~\ref{Sec_MU}. Figure~\ref{Fig_FinalObsResult}b summarises the results of the reference position measurements \cite{Pollinger2026}. The black square  symbols represent the change of the mirror position $\Delta l_{\rm drift}$ at the time of the respective measurement cycle with respect to the initial reference position. The magnitude of the observed drifts achieves up to \qty{0.2}{\milli\metre}. The value is correlated with the daytime during and thus, temperature at the time of the measurement. The remnant thermal expansion of the system hence seems to dominate the observed mechanical drift of the reference point while smaller deviations on shorter time scales can be explained by minor manual interferometer optics realignments in between consecutive observations. To compensate for mechanical drifts, the change in reference position $\Delta l_{\rm drift}$ is subtracted from the absolute length obtained by the SWL measurements. Thus, all measurement are projected to the reference point at the time of the initial measurement. The red circles in Fig.~\ref{Fig_FinalObsResult}b indicate the uncertainty of the mirror position as absolute value of the difference between mirror measurements before and after the SWL measurement. This remnant position uncertainty remains below $\qty{0.011}{\milli\metre}$, about a factor of 20 smaller than the correction. The coarse lengths values are not drift--corrected as there is no dedicated mirror position measurement performed for speed reasons.  On the same day as the interferometric measurements, the baselines were also measured by a commercial electronic distance meter (Leica TDA5005) using the same mechanical mounts. The obtained measurement results, $l_{1, {\rm ref}}$ = \qty{28004.92(0.12)}{\milli\metre} for the short and $l_{2, {\rm ref}}$ = \qty{224002.68(0.92)}{\milli\metre} for the long baseline serve as reference values in this discussion. Since there is no independent offset calibration of the TeleYAG-II interferometer available for these measurements, the deviation of the first absolute distance measurement at \qty{28}{\metre} of $l_{\rm off} = \qty{652.371}{\milli\metre}$ from the reference value is considered as instrument offset  and subtracted from all interferometric measurement results. This sets the deviation  $l_{\rm ref} - l_{\rm TeleYAG-II}$ of the first \qty{28}{\metre} measurement to 0. The resulting offset-corrected deviations of all measurement median values are depicted in Fig.~\ref{Fig_FinalObsResult}a. The uncertainty of the reference value are derived from the manufacturer specifications. Assuming a coverage factor $k$ of 2, the corresponding uncertainty is indicated by the grey area in the background of  Fig.~\ref{Fig_FinalObsResult}a. After the instrument offset correction, all high resolution measurement results fit within \qty{0.25}{\milli\metre} to the reference value, very well within the uncertainty of the reference measurement. The best estimate of the two baseline distances can be derived from the weighted means $\tilde{l}_{1}$ = \qty{28004.914(0.069)}{\milli\metre} and $\tilde{l}_{2}$ = \qty{224002.866(0.032)}{\milli\metre} of the various individual measurement results, the uncertainties again being given for a coverage factor $k$ of 2. The weighted mean values are indicated by green dashed lines in Fig.~\ref{Fig_FinalObsResult}a. Although this value neglects by construction the uncertainty of the instrument offset, it remains remarkable that the uncertainty of these weighted means is well below \qty{0.1}{\milli\metre}. The uncertainty of the longer distance is by a factor of 2.2 smaller than the uncertainty of the shorter one. This value is pretty close to the theoretical scaling of the number of measurements entering the weighted mean value $\sqrt{11}/\sqrt{2} = 2.3$.  Although this should not be over-interpreted, it should be noted that this scaling implies that the uncertainty of the individual measurements is comparable for both distances, i.e. does not significantly changes for a distance of \qty{28}{\metre} to a distance of \qty{224}{\metre}.

\section{Measurement Uncertainty}\label{Sec_MU}
In this section, the measurement uncertainty of the experimental results presented in Sec.~\ref{Sec_Results} is discussed following the guide to the expression of uncertainty in measurement (GUM) \cite{BIPM}. To assess the GUM type A measurement uncertainty, the experimental standard deviation is discussed first.  The raw data taken with \qty{3.8125}{\kilo\hertz} show an experimental standard deviation $\sigma$  in the order of \qty{80}{\micro\meter}. There are several sources of origin. For once, there are inevitable mechanical instabilities of the complete setup, including interferometer head, target setup, but also the baseline pillars. Furthermore, the OPLL is not perfect. For continuous fixed frequency observations,  the control loop keeps the frequency difference $\Delta f$ only within a bandwidth of typically \qty{150}{\hertz} over \qty{1}{\second} to the target value. The resulting variation of the scale is not compensated in these experiments and hence leads to a statistical component of the length deviation.  In addition, air turbulence can be expected to have a statistical impact on the measurement result. According to the results of Sec.~\ref{Sec_Results_RawData}, averaging the observation data  can  significantly reduce the impact of the type A uncertainty on the final result.  Applying a \qty{5.4}{\second}-bin-average, the remnant vibrations and air turbulence effects lead in average to uncertainty contributions of $\sigma^{\rm d}(l_1) = \qty{11.4}{\micro\meter}$ for the shorter distance $l_1 = \qty{28}{\meter}$ during daytime (d). In case of the longer distance $l_2 = \qty{224}{\meter}$  an average standard deviation of $\sigma^{\rm d}(l_2) = \qty{22.2}{\micro\meter}$ during daytime and of $\sigma^{\rm n}(l_2) = \qty{13.4}{\micro\meter}$ during nighttime (n) are observed. 

Besides statistical errors, several systematic sources of uncertainty are to be taken into account. At first, uncertainty contributions by the interferometric measurement principle are discussed here. Most critical for the traceability to the SI definition of the metre is the uncertainty of the scale.  Equation~(\ref{Eq_Def_Synthetic_Wavelength_by_frequency}) ties the system scale to the OPLL frequency difference $\Delta f$ which is derived from a single master clock (see Fig.~\ref{Fig_concept_PSS}a). The deviation from the target frequency has been discussed as type A  uncertainty contribution, but the absolute value of the reference frequency implies a systematic uncertainty contribution, too. In the field, we use an older rubidium \qty{10}{\mega\hertz} frequency standard (Scientific Instruments PRS10) due to its compactness and robustness. In May 2024, its frequency was determined to $f_{\rm ref} = \qty{10000001.66}{\hertz}$ against an SI-traceable reference clock of higher quality. The manufacturer specifies the systems frequency stability to be better than $5\times10^{-10}$ per year \cite{RbSpecs}. Hence, a relative uncertainty of $\frac{u(f_{\rm ref})}{f_{\rm ref}} = 1\times10^{-9}$ for the frequency at the time of the  experiment can be assumed. For the purpose of the discussion here, we consider the uncertainty contribution of the OPLL to the laser frequency $\Delta f$  negligible. Then, the frequency difference $\Delta f$ can be approximated as a multiple $a \in \mathbb{R}$ of the reference frequency $f_{\rm ref}$, i.e. $\Delta f = a f_{\rm ref}$. According to Eq.~(\ref{Eq_LengthFrom2Colour}) and (\ref{Eq_Def_Synthetic_Wavelength_by_frequency}), the measured length of a synthetic wavelength measurement in the TeleYAG-II is directly proportional to the laser frequency difference $\Delta f$. The uncertainty contribution to the synthetic length measurement by the frequency uncertainty $u^{\rm ref}(l)$ hence scales with the baseline length $l$ according to $u^{\rm ref}_{\rm swl}(l) =  \frac{u(f_{\rm ref})}{f_{\rm ref}} \times l = 10^{-9}\times l$. Using the equivalent approximation $\Delta f_1  = a_1 f_{\rm ref}$ and $\Delta f_2  = a_2 f_{\rm ref}$ also for the PSS length solution given by Eq.~(\ref{Eq_Def_PSS_Wavelength_From_Delta_fs}), it can be shown that the dependence of the uncertainty of the pseudo-supersynthetic length similarly depends on the uncertainty of the reference frequency by $u^{\rm ref}_{\rm pss}(l) = 10^{-9}\times l$. For the longer distance of $l_2 = \qty{224}{\metre}$, this corresponds to an uncertainty contribution of $u^{\rm ref}(l_2) = \qty{0.23}{\micro\meter}$, for the shorter distance of $l_1 = \qty{28}{\metre}$ to only $u^{\rm ref}(l_1) = \qty{0.029}{\micro\meter}$. 

Further to the interferometric  measurement principle, the core assumption for the geometric interpretation of the interferometric phase given by Eq.~(\ref{Eq_Interferometric_Phase_basic}) must be critically reviewed.  This equation is strictly valid only for interfering plane waves. The TeleYAG-II interferometer does not emit plane waves, but Gaussian beams with pronounced curvatures over long distances. In this case, the measured interferometric phase is modified  \cite[chap.~9.3.1.2]{Andreas2018}: 
\begin{equation}
 \phi(l, \lambda)=4\pi \frac{n^{\rm ph}(\lambda) l}{\lambda} - \arctan\left(\frac{2l-z_{\rm R}(\lambda)}{2z_{\rm R}(\lambda)}\right) \label{Eq_Phase_Gaussian_Beam_Interference}
\end{equation} 
with the Rayleigh length $z_{{\rm R}}(\lambda) = \omega_0^2 \pi n_{{\rm Ph}}(\lambda)\lambda^{-1}$. Inserting (\ref{Eq_Phase_Gaussian_Beam_Interference}) into Eq.~(\ref{Eq_SynPhase}) and (\ref{Eq_Length_From_Synthetic_WL}) it is clear that the systematic error $\delta l^{\phi}$ due to this approximation for the length measured with the synthetic wavelength length can be expressed by
\begin{equation}
\delta l^{\phi}(l)=\frac{\Lambda^{\rm swl}}{4\pi n^{\rm 2WL}} \left( \arctan(\frac{2l-z_{\rm R}(\lambda_1)}{2z_{\rm R}(\lambda_1)}) - \arctan(\frac{2l-z_{\rm R}(\lambda_2)}{2z_{\rm R}(\lambda_2)})\right).
\end{equation}
For the distances in this study, this systematic error remains below $u^{\phi} (l)= \qty{0.006}{\micro\metre}$.
As a third interferometric uncertainty contribution, we need to consider the intrinsic phase resolution of the phase meter. For the TeleYAG-II system it is given by $\delta \phi^{\rm res} =  \qty{0.13}{\milli \radian}$ (see Sec.~\ref{Sec_Method_Signal_Demodulation}). This implies a standard uncertainty of the synthetic phase $u^{\rm res} (\phi)= \sqrt{2}/(2\sqrt{3}) \times \delta \phi^{\rm res} = \qty{0.053}{\milli \radian}$, corresponding to an uncertainty of the synthetic length result of $u^{\rm res}(l) = \qty{0.030}{\micro\meter}$. Finally, there is the impact of interferometric non-linear effects. If the interferometer alignment is carefully optimised for the operating wavelengths at \qty{532}{\nano\meter}, cyclic errors cannot be resolved in the raw data of dedicated characterisation experiments  and are therefore considered negligible in this context. 

\begin{table}
\caption{Uncertainty contributions related to the uncertainty of the two-colour group refractive index $n^{\rm 2WL}$. Average values observed during the campaign are shown for the different working distances $l_1 = \qty{28}{\metre}$  and $l_2 = \qty{224}{\metre}$, grouped for day (d) and night (n) measurements. The sensitivity coefficients $c_{X_i}(l)  $ are given for the working wavelength duplet at $(\lambda_1, \lambda_2)$ = $(\qty{532.238}{\nano\metre}, \qty{532.198}{\nano\metre})$ as averaged for all measurements to indicate their order of magnitude.  }\centering
\begin{tabular}{| l | c | c| c | c|c | c |c|}
\hline
					& \multicolumn{3}{c|}{Standard Uncertainty} 														&Sensitivity 							& \multicolumn{3}{c|}{Uncertainty}\\
					& \multicolumn{3}{c|}{} 																			&in \qty{}{m^{-1}} 						& \multicolumn{3}{c|}{in \qty{}{\micro\metre}}  \\
$X_i$				&$u^{\rm d}_{l_1}(X_i)$	&$u^{\rm d}_{l_2}(X_i)$		&$u^{\rm n}_{l_2}(X_i)$		&$c_{X_i}(l)  $							&$u^{\rm d}_{l_1}(l)$	&$u^{\rm d}_{l_2}(l)$	&$u^{\rm n}_{l_2}(l)$ \\
\hline
$\sigma_t(T)$			&\qty{31}{\milli\kelvin}			&\qty{0.13}{\kelvin}				&\qty{34}{\milli\kelvin}				& \qty{-1.02e-6}{\kelvin^{-1}}			&\qty{0.91}{}				&\qty{30}{} 					&\qty{7.7}{} \\
$\Delta T_{\rm long}$	&\qty{94}{\milli\kelvin}			&\qty{68}{\milli\kelvin} 				&\qty{45}{\milli\kelvin} 				& \qty{-1.02e-6}{\kelvin^{-1}}			&\qty{2.8}{}				&\qty{15}{} 					&\qty{10}{} \\
$\Delta T_{\rm vert}$	&\qty{11}{\milli\kelvin}			&\qty{7.4}{\milli\kelvin}				&\qty{8.0}{\milli\kelvin}				&\qty{-1.02e-6}{\kelvin^{-1}}			&\qty{0.32}{}				&\qty{1.6}{} 					&\qty{1.8}{}\\
$\Delta T_{\rm sys}$	&\qty{40}{\milli\kelvin}			&\qty{40}{\milli\kelvin}				&\qty{40}{\milli\kelvin}				&\qty{-1.02e-6}{\kelvin^{-1}}			&\qty{1.2}{}				&\qty{8.9}{} 					&\qty{9.2}{} \\
$\sigma_t(p)$			&\qty{2.9}{\pascal}				&\qty{5.5}{\pascal}					&\qty{3.9}{\pascal}					&\qty{2.87e-5}{\pascal^{-1}}				&\qty{0.24}{}				&\qty{3.5}{} 					&\qty{2.5}{}\\
$p_{\rm cal}$ 			&\qty{1.5}{\pascal}				&\qty{1.5}{\pascal}					&\qty{1.5}{\pascal}					&\qty{2.87e-5}{\pascal^{-1}}				&\qty{0.12}{}				&\qty{0.96}{} 					&\qty{0.97}{}\\
$\Delta p$ 			&\qty{10}{\pascal}				&\qty{10}{\pascal}					&\qty{10}{\pascal}					&\qty{2.87e-5}{\pascal^{-1}}				&\qty{0.82}{}				&\qty{6.4}{} 					&\qty{6.5}{}\\
$\sigma_t(RH)$		&\qty{0.4}{\percent}${\rm RH}$	&\qty{1.7}{\percent}${\rm RH}$		&\qty{0.1}{\percent}${\rm RH}$		&\qty{-4.81e-9}{\percent}${\rm RH^{-1}}$	&\qty{0.1}{}				&\qty{2.3}{} 					&\qty{0.0}{}\\
$RH_{\rm cal}$		&\qty{0.9}{\percent}${\rm RH}$	&\qty{0.9}{\percent}${\rm RH}$		&\qty{0.9}{\percent}${\rm RH}$		&\qty{-4.81e-9}{\percent}${\rm RH^{-1}}$	&\qty{0.11}{}				&\qty{1.1}{}					&\qty{0.97}{}\\
$\Delta RH$			&\qty{2.0}{\percent}${\rm RH}$	&\qty{2.0}{\percent}${\rm RH}$		&\qty{2.0}{\percent}${\rm RH}$		&\qty{-4.81e-9}{\percent}${\rm RH^{-1}}$	&\qty{0.24}{}				&\qty{2.4}{} 					&\qty{2.2}{}\\
$x_c$ 				&\qty{58}{ppm}				&\qty{58}{ppm}					&\qty{58}{ppm}					&\qty{1.49e-10}{ppm^{-1}}				&\qty{0.24}{}				&\qty{2.0}{}					&\qty{1.9}{}\\
\hline
\multicolumn{5}{|r|}{Total environmental uncertainty contribution $u^{\rm env}(l)$:}																								&\qty{3.3}{}				&\qty{36}{} 					&\qty{18}{}\\
\hline
\end{tabular}
\label{Tab_EnvUncertainty}
\end{table}

Moreover, a core problem for an outdoor optical measurement is the determination of the index of refraction. At the WUT200 baseline, a weather station designed for distances of a few hundred metres is used. It consists of five Pt-100 temperature sensors at beam height distributed along the complete baseline. They provide the average temperature $T_{\rm av}$  which varied between \qty{13.2}{\celsius} and \qty{18.5}{\celsius} during the measurement campaign. They also allow an estimate of the longitudinal gradient $\Delta T_{\rm long}$. It is approximated by half the spread of the five sensors with a uniform probability distribution function (PDF).  Two additional temperature sensors measure the vertical temperature gradient $\Delta T_{\rm vert}$. The sensors are shielded for radiation, but not equipped with active ventilation. Their possible systematic deviation $\Delta T_{\rm sys}$ including the calibration uncertainty can conservatively be estimated to approximately \qty{40}{\milli\kelvin} \cite{Pollinger2012}. The measurements discussed here are performed at an ambient pressure $p$ ranging between \qty{988.4}{\hecto\pascal} and \qty{990.9}{\hecto\pascal}. A single sensor (Setra 370) is used with a calibration uncertainty contribution $p_{\rm cal}$ of \qty{0.15}{\hecto\pascal}. An uncertainty due to variance of the ambient pressure $\Delta p$ for the \qty{224}{\metre}--distance of \qty{0.10}{\hecto\pascal} appears reasonable \cite{Pollinger2012}. The relative humidity $RH$ varies between \qty{51.5}{\percent}${\rm RH}$ and \qty{91.8}{\percent}${\rm RH}$ during the measurements. It is measured by a single instrument (Testo 650) with a calibration uncertainty $u_{\rm cal}(RH)$ of \qty{0.9}{\percent}${\rm RH}$. A homogeneity $\Delta RH$ of \qty{2.0}{\percent}${\rm RH}$ is assumed based on literature values for the conditions during the measurement \cite{Pollinger2012}. For all observables $T$, $p$, and $RH$, the standard deviation of the effective value over time $\sigma_t$ can be derived from the measurement data. They are suitable indications for the uncertainty contribution due to the imperfect temporal resolution of the data sampling. Typical values are included in Table~\ref{Tab_EnvUncertainty}. The CO$_2$ contents $x_c$ is finally estimated to \qty{450}{ppm} with a standard uncertainty of $100/\sqrt{3}\,{\rm ppm}$. The sensitivity coefficient $c_{X_i}$ for each environmental uncertainty contribution $X_i$ is given by the partial derivative of the two-colour group index of refraction $n^{\rm 2WL}$ defined in Eq.~(\ref{Eq_Two_colour_index_of_refraction}) with respect to the respective quantity $X_i$:
\begin{equation}
c_{X_i} = \frac{\partial n^{\rm 2WL}(\lambda_1, \lambda_2, T, p, RH, x_c)}{\partial X_i}. \label{Eq_sens_env}
\end{equation}
Table \ref{Tab_EnvUncertainty} compiles mean environmental uncertainty contributions $X_i$ for the TeleYAG-II parameters during the measurement campaign. The sensitivity coefficients $c_{X_i}$ depend on the specific ambient conditions at the time of the measurement. In Table~\ref{Tab_EnvUncertainty},  the mean value of all measurements is depicted to provide an impression of the order of magnitude.
Average values for day (d) and night (n) measurements are grouped for the $l_2 = \qty{224}{\metre}$ measurement. Evidently, conditions are more unstable during the day than at night for the measurement discussed here. The influence of changing conditions during the measurement, coupled with greater local heterogeneity, results in environmental conditions contributing  twice as much to the uncertainty during the day as at night ($u^{\rm d}_{l_2} = \qty{36}{\micro\metre}$ vs. $u^{\rm n}_{l_2} = \qty{18}{\micro\metre}$). The uncertainty contributions of the index of refraction scale with the measurement length. The ratio of the two uncertainties of the daytime measurements $u^{\rm d}_{l_2} / u^{\rm d}_{l_1}$ is approximately 11, slightly larger than the corresponding length ratio $l_2 / l_1$ of 8. The environmental conditions during the shorter baseline measurements were obviously slightly more stable than during the longer baseline measurements. For the uncertainty bars in Fig.~\ref{Fig_FinalObsResult} they have been individually evaluated. In this context, it should be noted that the environmentally-induced uncertainty contributions for the PSS measurement results are calculated in complete analogy, replacing the two-wavelength index of refraction in Eq.~(\ref{Eq_sens_env}) by the PSS index of refraction defined by (\ref{Eq_PSS_index_of_refraction}). They have been considered for the data shown in Fig.~\ref{Fig_Sweep}b.

Finally, the mechanical influence of the setup on the measurement uncertainty needs to be considered. Despite all constructive measures, mechanical drifts of the setup cannot fully be prevented due to realignments, remnant temperature sensitivities and other effects. To compensate such drifts, the retractable mirror (cf. Fig. \ref{Fig_setup_TeleYAG_II}) is used as an internal reference point. The red dots in Fig.~\ref{Fig_FinalObsResult}b show the magnitude of these short-term drifts $\delta \Delta l _{\rm drift}$. They are smaller than $\qty{11}{\micro\meter}$. They are determined for each observation separately, but a meaningful compensation cannot be applied. So we use this quantity as an estimate for the uncertainty of the drift compensation. Assuming a uniform PDF, the drift uncertainty contribution can hence be concluded as $u^{\rm drift} (l) = \qty{6.4}{\micro\meter}$. Another issue to be taken into account is the violation of the Abbe principle in such a baseline distance measurement configuration. The working height over the tribrach (and thus over the distance to be measured) for TeleYAG-II and the reflector is $h_{\rm beam} = \qty{0.301}{\metre}$. The systems are levelled by a high precision water level (Leica SNLL121) with a nominal uncertainty of $\delta \theta = \qty{3.8}{\milli\radian}$. This corresponds to a possible deviation from the reference point of $\Delta l_{\rm level} = h_{\rm beam} *\sin\left( \delta\theta \right) = \qty{44}{\micro\metre}$. Since both instrument and reflector are levelled, and a uniform PDF can be assumed, the uncertainty contribution of the levelling $u^{\rm level}(l)$ amounts to $u_{\rm level}(l) = \sqrt{2} \Delta  l_{\rm level} / \sqrt{3} = \qty{36}{\micro\metre}$. In between the short and the long-range measurement, the instrument is rotated by \qty{180}{\degree}. The deviation of the instrument's rotational invariant point from the beam axis  \cite{Guillory2023} was carefully aligned in the manufacturing process by a coordinate measurement machine to be remain below $u^{\rm inv} (l)= \qty{30}{\micro\metre}$. Finally, it should be noted that centring uncertainties due to an imperfect tribrach are irrelevant for the comparison result as the reference instrument was mounted on the same tribrach as the TeleYAG-II interferometer. However, for a true absolute distance measurement, these should be explicitly considered \cite{Guillory2023}.

\begin{table}
\centering
\begin{threeparttable}
\caption{Meausurement uncertainty budget for typical measurement results in the campaign for the different working distances $l_1 = \qty{28}{\metre}$  and $l_2 = \qty{224}{\metre}$.  }
\begin{tabular}{| l | c | c| c|}

\hline
Quantity										& \multicolumn{3}{c|}{Uncertainty contribution}  \\
											& \multicolumn{3}{c|}{in \qty{}{\micro\meter}}  \\
$X_i$										&$u^{\rm d}_{l_1}(l) $	&$u^{\rm d}_{l_2}(l)$	& $u^{\rm n}_{l_2}(l)$ \\ 
\hline
Experimental standard deviation $\sigma$ \tnote{a}	&11.4 				&22.2				&13.4\\
Environment $u^{\rm env}(l)$					&3.3					&36					&18\\
Reference frequency $u^{\rm ref}(l)$				&0.029				&0.23				&0.23\\
Gaussian beam phase shift $u^{\phi}(l)$			&0.006				&0.006				&0.006	\\
Intrinsic phase meter resolution $u^{\rm res}(l)$		&0.030				&0.030				&0.030\\
Short-term drifts $u^{\rm drift}(l)$				&6.4					&6.4					&6.4	\\
Levelling $u^{\rm level}(l)$						&36					&36					&36\\
Invariant point $u^{\rm inv}$				&30					&30					&30\\
\hline
combined measurement uncertainty ($k$ = 1):		&49					&\qty{63}{}			&\qty{52}{}\\
expanded measurement uncertainty ($k$ = 2):		&98					&\qty{1.3e2}{} 		&\qty{1.0e2}{}\\
\hline
\end{tabular}
\begin{tablenotes}
\item[a] \qty{5.4}{\second} averaging
\end{tablenotes}
\end{threeparttable}
\label{Tab_MUB_global}
\end{table}

Table~\ref{Tab_MUB_global} compiles all discussed uncertainty contributions for the full measurement uncertainty budget. Summed up according to GUM rules, the expanded measurement uncertainty varies between \qty{98}{\micro\metre} and \qty{1.3e2}{\micro\meter} for the two distances and measurement conditions analysed in this study for a coverage factor $k = 2$. It should be noted that for the calm environmental conditions experienced during the night measurement the measurement uncertainty for short and long length $l_1$ and $l_2$ are almost identical. The comparison of experimental measurement results with expanded measurement uncertainties to the reference values in Fig.~\ref{Fig_FinalObsResult}a shows a satisfactory coincidence. Although limited by the relatively high uncertainty of the reference values, this result supports the derived measurement uncertainty.

\section{Discussion and Conclusions}\label{Sec_Discussion}
The overall target of this study is the development of a reliable, primary standard for the calibration of geodetic baselines. Interferometry allows in principle the implementation of the SI definition of the metre with the lowest uncertainty. But unlike many other studies in the past, the TeleYAG-II interferometer is supposed to go beyond proof-of-principle stage for the realisation of such an absolute interferometer to real-life capability.  Laser source, mechanics and optics as well as data collection have been designed for reliable continuous application in the field. The all-fibred optical source and fibre-moderated beam superposition enables an efficient and robust optics alignment on site. This leads to a very substantial reduction of the realignment need of the interferometer head compared to the original TeleYAG interferometer.  Continuous interference datasets over more than two hundred metres in the field  can be taken over tens of minutes without substantial realignments or additional a-posteriori data filtering. Allan deviations and experimental standard deviations show almost neglectable impact of the length on the uncertainty for the ranges investigated in this study. This is, however, also aided by the exceptionally calm weather conditions during the measurement campaign. The introduction of an artificial intrinsic reference point  reduces the impact of long-term remnant drifts below \qty{10}{\micro\metre} and allows long-term calibration of the system. It should be noted, however, that the NIR data have not achieved this quality level yet. Thus, we have decided not to present them in this paper. There are more pronounced non-linearities in the NIR case as well as chromatic imaging problems with the telescope optics. We have performed intense investigations into this and are currently working on technical solutions. When these issues are resolved, the refractive index compensation can be performed using the intrinsic dispersive correction or two-colour method \cite{Bender1965, MeinersHagen2008} instead of the complex sensor network weather station. Such a true-real time compensation will remove the Nyquist limit of the environmental sampling. The increase of the Allan deviation for the long-distance measurements observed in Fig.~\ref{Fig_RawData}b can be expected to be substantially reduced. This is of particular importance for the measurement of even longer baselines.

The performance in the VIS wavelength regime, however, is remarkable. The key for this result is the new concept of TSWL and finally PSS interferometry. It achieves the accuracy of classical SWL interferometry, without the need of multiple laser sources for the realisation of the SWL chain. It allows a true NAR over an almost indefinite distance. Quasi-unlimited PSS wavelengths can be easily constructed by subsequent measurements of similar SWLs. Figure~\ref{Fig_Sweep}c shows the textbook--style unwrapping chain that can be constructed by this method. In future measurements we will further refine the frequency choice to avoid duplicates present in the data of Fig.~\ref{Fig_Sweep}. Classical FSI or VSW interferometry allow in principle also an unambiguous indefinite absolute distance measurement. But ultimately, they run in the problem that full phase cycles must be tracked during the frequency variation as the full unwrapped phase is required for the correct interpretation of the length. The imperfect closure of the phase sweeping circle in Fig.~\ref{Fig_Sweep}a shows the impact of these cycle slips. The realisation of the scale by an OPLL of variable external reference frequency does not only enable direct traceability to the SI definition of the metre, but also enables the exploitation of a new concept for absolute distance interferometry. The final uncertainty of the TSWL approach is effectively determined by the length of the shortest synthetic. In this case, this is given by the bandwidth of the Nd:YAG lasers used. To reduce the shortest synthetic further for other applications, other optical sources must be considered. But for the geodetic application intended by the TeleYAG-II instrument with its accuracy limit ultimately dictated by the reproducibility of the surveying tribrach  in the order of \qty{35}{\micro\meter}, the bandwidth of the chosen Nd:YAG configuration is sufficient. It should be noted, however, that the TSWL and PSS measurement principle are highly sensitive to distance variations during the frequency variation measurement. Fortunately, measurement speed is not a primary issue of concern in case of the calibration of geodetic baselines since they are built for long-term stability.

We must concede that the study presented here is not fully sufficient to establish the TeleYAG-II interferometer as primary standard yet. While the comparison of the experimental results and their derived measurement uncertainties are in agreement with the reference values, verification measurements over more distances with substantially lower uncertainties is desirable for a final performance evidence. More importantly, the long-term stability of the reference point of the system must be studied in more detail. In particular, a calibration procedure for the instrument offset must be developed. Standard two-face measurement methods used in geodesy, which involve measurements in the first and second instrument positions, require the telescope to be rotated vertically by \qty{180}{\degree}. This is not feasible with the sandwich configuration of the TeleYAG-II interferometer.

In conclusion, the study shows the validity of the design concepts behind the TeleYAG-II distance meter and gives a first impression of its capabilities. The spatial separation of optical source, interferometer head and detector unit limit the exposure of optical components to the environment to the inevitable components only.  The use of PM-PCFs as wide-range wavelength division multiplexer and the careful mechanical design of the interferometer head enables a continuous operation of the system in the typical environment for a baseline calibration. After resolving the remaining issues with the NIR wavelength, the impact of the refractivity-compensation over longer distances will be further reduced. But the achieved maximum expanded measurement uncertainty \qty{0.13}{\milli\metre} for a distance of \qty{224}{\metre} is already remarkable. While the prospect of a robust, primary standard for geodetic baseline is in itself of high value for length metrology and geodesy, the new PSS concept introduced in this paper is of broader relevance. It requires a frequency-tunable source and a high-accuracy measurement of  the SWL frequency difference at the time of the measurement. Furthermore, the target position should not rapidly change beyond the tracking window of an interferometric system. If these conditions are fulfilled, the combination of TSWL and PSS allows robust and cost-efficient  absolute distance interferometry with an unlimited NAR.

\ack{The authors would like to thank Max Peglow, Uwe Langner, Daniel Fiala, Martin Wedde, Lea-Jean Fr\"{o}mel  and Patrik Knigge für technical support and Ren\'{e} Sch\"odel and Christoph Weichert for fruitful discussions. The authors are grateful to Jakob Silbermann for his constructive design of the achromatic expansion optics. They would also like to thank Kinga W\k{e}zka, Mariusz Wi\'{s}niewski, Marco Pisani, Jeremiah Sepp\"a,  and Massimo Zucco for the good collaboration during the WUT200 measurement campaign. Kinga W\k{e}zka also provided the external reference value. Finally, they would also like to thank Frank Kahlecke, Diether Ohland and Frank R\"{u}tters for the safe transport of the instrument to various locations.}

\funding{The projects 18SIB01 GeoMetre and 20IND02 DynaMITE have received funding from the EMPIR programme cofinanced by the Participating States and from the European Union’s Horizon 2020 research and innovation programme. F.Po., H.Y. and P.K. acknowledge partial funding by the Bundesministerium für Wirtschaft und  Energie (TransMET 2024-I-03).}

\roles{F.Po.: Conceptualisation (interferometer, optical source, complete experiment), Methodology (PSS, TSWL, SWL), Software (data analysis and post-processing), Validation, Formal Analysis, Investigation, Resources (interferometer alignment, weather station), Data Curation, Visualisation, Project administration, Funding acquisition, Supervision, Writing - original draft.  F.Pi.: Conceptualisation (interferometer head), Resources (interferometer head mechanics), Writing - review \& editing. T.M.: Resources (initial interferometer head alignment, weather station),  Investigation (initial measurement campaigns), Writing - review \& editing. H.Y.: Investigation (nonlinearities), Resources, Validation, Writing - review \& editing. A.S.: Methodology (PSS, measurement uncertainty analysis),  Writing - review \& editing.  G.P.: Conceptualisation (optical source), Resources (optical source), Validation (nonlinearities), Writing - review \& editing.  S.W.: Conceptualisation (interferometer alignment), Resources (interferimeter alignment unit), Writing - review \& editing. FS: Conceptualisation (interferometer head), Writing - review \& editing. P.K.: Conceptualisation (OPLL, phase meter, optical source), Methodology  (PSS, TSWL, SWL), Software (data acquisition), Validation, Investigation, Resources (OPLL, phase meter, weather station), Supervision, Visualisation, Writing - original draft (Sec. 3.1), Writing - review \& editing. }

\data{The data that support the findings of this study are openly available at the following DOI: \newline \href{https://doi.org/10.7795/720.20261014} {10.7795/720.20261014}.}


\bibliographystyle{iopart-num}
\bibliography{TeleYAG-II}

\providecommand{\newblock}{}
\begin{thebibliography}{10}
\expandafter\ifx\csname url\endcsname\relax
  \def\url#1{{\tt #1}}\fi
\expandafter\ifx\csname urlprefix\endcsname\relax\def\urlprefix{URL }\fi
\providecommand{\eprint}[2][]{\url{#2}}

\bibitem{unece1996land}
{United Nations Economic Commission for Europe (UNECE)} 1996 Land
  administration guidelines Tech. Rep. E.96.II.E.7 United Nations New York and
  Geneva
  \urlprefix\url{https://unece.org/fileadmin/DAM/hlm/archive/landadmin.guidelines.pdf}

\bibitem{Niemeier2008}
Niemeier W 2008 {\em Ausgleichungsrechnung: Statistische Auswertemethoden in
  der Geodäsie\/} 2nd ed (Walter de Gruyter)

\bibitem{Gross2009}
Gross R, Beutler G and Plag H~P 2009 Integrated scientific and societal user
  requirements and functional specifications for the {GGOS} {\em Global
  Geodetic Observing System Meeting the Requirements of a Global Society on a
  Changing Planet in 2020\/} ed Plag H~P and Pearlman M (Berlin: Springer
  Verlag) pp 209--224

\bibitem{Baird1968}
Baird K~M 1968 {\em Metrologia\/} {\bf 4} 135
  \urlprefix\url{https://doi.org/10.1088/0026-1394/4/3/007}

\bibitem{Rueeger1996}
R\"{u}eger J~M 1996 {\em Electronic Distance Measurement\/} (Springer--Verlag,
  Berlin)

\bibitem{Pollinger2012}
Pollinger F, Meyer T, Beyer J, Doloca N~R, Schellin W, Niemeier W, Jokela J,
  Häkli P, Abou-Zeid A and Meiners-Hagen K 2012 {\em Measurement Science and
  Technology\/} {\bf 23} 094018
  \urlprefix\url{https://doi.org/10.1088/0957-0233/23/9/094018}

\bibitem{Wu2017}
Wu H, Zhao T, Wang Z, Zhang K, Xue B, Li J, He M and Qu X 2017 {\em Applied
  Physics Letters\/} {\bf 111} 251901 ISSN 0003-6951
  \urlprefix\url{https://doi.org/10.1063/1.4999537}

\bibitem{Wisniewski2026}
Wiśniewski M, Lubowicki T, Wezka K, Szpunar R, Próchniewicz D, Guillory J,
  Truong D, Wallerand J~P, Baselga S, García-Asenjo L, Garrigues P and Luján
  R 2026 {\em Measurement\/} {\bf 275} 121347 ISSN 0263-2241
  \urlprefix\url{https://doi.org/10.1016/j.measurement.2026.121347}

\bibitem{Jokela2026}
Jokela J, H\"{a}kli P and Saari T 2026 Interference measurements at the nummela
  standard baseline in 2024 for maintenance of a world-class length measurement
  standard {\em Proceedings of the XXVIII FIG Congress, May 24–29, 2026, Cape
  Town, South Africa\/}
  \urlprefix\url{https://www.fig.net/resources/proceedings/fig_proceedings/fig2026/papers/ts07a/TS07A_jokela_haekli_et_al_13703.pdf}

\bibitem{Vaisala1923}
V\"{a}is\"{a}l\"{a} Y 1923 {\em Publications of Finish Geodetic Institute\/}
  {\bf 2} 1--22 \urlprefix\url{https://hdl.handle.net/10138/347240}

\bibitem{Baselga2022}
Baselga S, García-Asenjo L, Garrigues P and Luján R 2022 {\em Measurement
  Science and Technology\/} {\bf 33} 085020
  \urlprefix\url{https://doi.org/10.1088/1361-6501/ac6f45}

\bibitem{Guillory2016}
Guillory J, Šmíd R, García-Márquez J, Truong D, Alexandre C and Wallerand
  J~P 2016 {\em Review of Scientific Instruments\/} {\bf 87} 075105
  \urlprefix\url{https://doi.org/10.1063/1.4954180}

\bibitem{Guillory2019}
Guillory J, Teyssendier de~la Serve M, Truong D, Alexandre C and Wallerand J~P
  2019 {\em IEEE Transactions on Instrumentation and Measurement\/} {\bf 68}
  2260--2267 \urlprefix\url{https://doi.org/10.1109/TIM.2019.2902804}

\bibitem{Guillory2023}
Guillory J, Truong D, Wallerand J~P and Alexandre C 2023 {\em Measurement
  Science and Technology\/} {\bf 35} 025024
  \urlprefix\url{https://doi.org/10.1088/1361-6501/ad0a22}

\bibitem{Minoshima2000}
Minoshima K and Matsumoto H 2000 {\em Appl. Opt.\/} {\bf 39} 5512--5517
  \urlprefix\url{https://doi.org/10.1364/AO.39.005512}

\bibitem{Xu2020}
Xu X, Zhang Z, Zhang H, Zhao H, Xia W, He M, Li J, Zhai J and Wu H 2020 {\em
  Opt. Express\/} {\bf 28} 4398--4411
  \urlprefix\url{https://doi.org/10.1364/OE.381504}

\bibitem{Kim2023}
Kim W, Yang J, Jang J, Oh J~S, Han S, Kim S, Jang H, Kim Y~J and Kim S~W 2023
  {\em Sensors\/} {\bf 23} ISSN 1424-8220
  \urlprefix\url{https://doi.org/10.3390/s23218949}

\bibitem{Jiang2025}
Jiang Y, Liu Y, Li J, Zhu L, Xu T, Miao D, He M and Chang X 2025 {\em Opt.
  Express\/} {\bf 33} 36834--36847 ISSN 1094-4087
  \urlprefix\url{https://doi.org/10.1364/OE.573258}

\bibitem{MeinersHagen2015}
Meiners-Hagen K, Bosnjakovic A, Köchert P and Pollinger F 2015 {\em
  Measurement Science and Technology\/} {\bf 26} 084002
  \urlprefix\url{http://doi.org/10.1088/0957-0233/26/8/084002}

\bibitem{MeinersHagen2017}
Meiners-Hagen K, Meyer T, Mildner J and Pollinger F 2017 {\em Applied Physics
  Letters\/} {\bf 111} 191104 \urlprefix\url{https://doi.org/10.1063/1.5000569}

\bibitem{Pollinger2023}
Pollinger F, Baselga S, Courde C, Eschelbach C, Garc\'{i}a-Asenjo L, Garrigues
  P, Guillory J, Hedekvist P~O, Heloj\"{a}rvi T, Jokela J, Kallio U, Kl\"{u}gel
  T, K\"{o}chert P, L\"{o}sler M, Luj\'{a}n R, Meyer T, Neyezhmakov P, Pesce D,
  Pisani M, Poutanen M, Prellinger G, Sauthoff A, Sepp\"{a} J, Truong D,
  Underwood R, Wezka K, Wallerand J~P and Wi\'{s}niewski M 2023 {\em Applied
  Geomatics\/} {\bf 15} 371--381 ISSN 1866-928X
  \urlprefix\url{https://doi.org/10.1007/s12518-022-00487-3}

\bibitem{MeinersHagen2009}
Meiners-Hagen K, Schödel R, Pollinger F and Abou-Zeid A 2009 {\em Measurement
  Science Review\/} {\bf 9} 16--26
  \urlprefix\url{http://www.measurement.sk/2009/S1/p.html}

\bibitem{Falaggis2018}
Falaggis K, Ramirez-Andrade A~H, Towers D, Towers C and Porras-Aguilar R 2018
  {Multi-wavelength phase unwrapping: a versatile tool for extending the
  measurement range, breaking the Nyquist limit, and encrypting optical
  communications} {\em Interferometry XIX\/} vol 10749 ed Creath K, Burke J,
  Morris M~B~N and Davies A~D International Society for Optics and Photonics
  (SPIE) p 1074913 \urlprefix\url{https://doi.org/10.1117/12.2321987}

\bibitem{Benoit1898}
Beno\^{\i}t R 1898 {\em Journal de Physique Th\'{e}orique et Appliqu\'{e}e\/}
  {\bf 7} 57 \urlprefix\url{http://doi.org/10.1051/jphystap:01898007005700}

\bibitem{Michelson1894}
Michelson A~A 1894 {\em D\'{e}termination exp\'{e}rimentale de la valeur du
  m\`{e}tre en longueurs d'ondes lumineuses\/} (Paris: Gauthier-Villars et
  Fils, Impremieurs-libraires du Bureau International des Poids et M\'{e}sures)
  \urlprefix\url{http://catalogue.bnf.fr/ark:/12148/cb30943848p}

\bibitem{Wyant1971}
Wyant J~C 1971 {\em Appl. Opt.\/} {\bf 10} 2113--2118
  \urlprefix\url{http://doi.org/10.1364/AO.10.002113}

\bibitem{Groot1994}
de~Groot P~J 1994 {\em Appl. Opt.\/} {\bf 33} 5948--5953
  \urlprefix\url{http://doi.org/10.1364/AO.33.005948}

\bibitem{Pollinger2009}
Pollinger F, Meiners-Hagen K, Wedde M and Abou-Zeid A 2009 {\em Appl. Opt.\/}
  {\bf 48} 6188--6194 \urlprefix\url{https://doi.org/10.1364/AO.48.006188}

\bibitem{Bourdet1979}
Bourdet G~L and Orszag A~G 1979 {\em Appl. Opt.\/} {\bf 18} 225--227
  \urlprefix\url{https://doi.org/10.1364/AO.18.000225}

\bibitem{Matsumoto1984}
Matsumoto H 1984 {\em Appl. Opt.\/} {\bf 23} 973--974
  \urlprefix\url{https://doi.org/10.1364/AO.23.000973}

\bibitem{Yang2014}
Yang R, Pollinger F, Meiners-Hagen K, Tan J and Bosse H 2014 {\em Opt. Lett.\/}
  {\bf 39} 5834--5837 \urlprefix\url{https://doi.org/10.1364/OL.39.005834}

\bibitem{Yang2015}
Yang R, Pollinger F, Meiners-Hagen K, Krystek M, Tan J and Bosse H 2015 {\em
  Measurement Science and Technology\/} {\bf 26} 084001
  \urlprefix\url{https://doi.org/10.1088/0957-0233/26/8/084001}

\bibitem{Zhao2018}
Zhao X, Qu X, Zhang F, Zhao Y and Tang G 2018 {\em Opt. Lett.\/} {\bf 43}
  807--810 \urlprefix\url{https://doi.org/10.1364/OL.43.000807}

\bibitem{Park2024}
Park J, Jang Y~S and Jin J 2024 {\em Metrologia\/} {\bf 61}
  \urlprefix\url{https://doi.org/10.1088/1681-7575/ad1af1}

\bibitem{Li2025}
Li R, Tian H, Liu Y, Dong D, Sterczewski L~a and Zhou W 2025 {\em Opt.
  Express\/} {\bf 33} 10528--10537
  \urlprefix\url{https://doi.org/10.1364/OE.553690}

\bibitem{Yokoyama2009}
Yokoyama S, Yokoyama T, Hagihara Y, Araki T and Yasui T 2009 {\em Opt.
  Express\/} {\bf 17} 17324--17337
  \urlprefix\url{https://doi.org/10.1364/OE.17.017324}

\bibitem{Berg2012}
van~den Berg S~A, Persijn S~T, Kok G~J~P, Zeitouny M~G and Bhattacharya N 2012
  {\em Phys. Rev. Lett.\/} {\bf 108}(18) 183901
  \urlprefix\url{https://doi.org/10.1103/PhysRevLett.108.183901}

\bibitem{Zhu2018}
Zhu Z and Wu G 2018 {\em Engineering\/} {\bf 4} 772 -- 778
  \urlprefix\url{https://doi.org/10.1016/j.eng.2018.10.002}

\bibitem{Stone1999}
Stone J~A, Stejskal A and Howard L 1999 {\em Appl. Opt.\/} {\bf 38} 5981--5994
  \urlprefix\url{http://ao.osa.org/abstract.cfm?URI=ao-38-28-5981}

\bibitem{Kinder2002}
Kinder T and Salewski K~D 2002 {\em Journal of Optics A: Pure and Applied
  Optics\/} {\bf 4} S364--S368
  \urlprefix\url{https://doi.org/10.1088/1464-4258/4/6/381}

\bibitem{Dale2014}
Dale J, Hughes B, Lancaster A~J, Lewis A~J, Reichold A~J~H and Warden M~S 2014
  {\em Opt. Express\/} {\bf 22} 24869--24893
  \urlprefix\url{https://doi.org/10.1364/OE.22.024869}

\bibitem{Prellinger2015}
Prellinger G, Meiners-Hagen K and Pollinger F 2015 {\em Measurement Science and
  Technology\/} {\bf 26} 084003
  \urlprefix\url{https://doi.org/10.1088/0957-0233/26/8/084003}

\bibitem{Hartmann2008}
Hartmann L, Meiners-Hagen K and Abou-Zeid A 2008 {\em Measurement Science and
  Technology\/} {\bf 19} 045307
  \urlprefix\url{https://doi.org/10.1088/0957-0233/19/4/045307}

\bibitem{Floch2008}
Floch S~L, Salvadé Y, Mitouassiwou R and Favre P 2008 {\em Appl. Opt.\/} {\bf
  47} 3027--3031 \urlprefix\url{https://doi.org/10.1364/AO.47.003027}

\bibitem{Zhang2018}
Zhang S, Xu Z, Chen B, Yan L and Xie J 2018 {\em Opt. Express\/} {\bf 26}
  9273--9284 \urlprefix\url{https://doi.org/10.1364/OE.26.009273}

\bibitem{Gelmini1994}
Gelmini E, Minoni U and Docchio F 1994 {\em Opt. Lett.\/} {\bf 19} 213--215
  \urlprefix\url{https://doi.org/10.1364/OL.19.000213}

\bibitem{Gelmini1995}
Gelmini E, Minoni U and Docchio F 1995 {\em Review of Scientific Instruments\/}
  {\bf 66} 4073--4080 ISSN 0034-6748
  \urlprefix\url{https://doi.org/10.1063/1.1145419}

\bibitem{Azouigui2010}
Azouigui S, Badr T, Wallerand J~P, Himbert M, Salgado J and Juncar P 2010 {\em
  Review of Scientific Instruments\/} {\bf 81} 053112 (pages~6)
  \urlprefix\url{https://doi.org/10.1063/1.3428729}

\bibitem{Czarske1995}
Czarske J, Philipps R and Freitag I 1995 {\em Applied Physics B\/} {\bf 61}
  243--248 \urlprefix\url{https://doi.org/10.1007/BF01082042}

\bibitem{Sorin1990}
Sorin W, Donald D, Newton S and Nazarathy M 1990 {\em IEEE Photonics Technology
  Letters\/} {\bf 2} 902--904 \urlprefix\url{https://doi.org/10.1109/68.62024}

\bibitem{Pilarski2021}
Pilarski F, Schmaljohann F, Weinrich S, Huismann J, Truong D, Meyer T,
  K\"{o}chert P, Sch\"{o}del R and Pollinger F 2021 Design and manufacture of a
  reference interferometer for long-range distancemetrology {\em Conference
  Proceedings - 21st International Conference \& Exhibition, Monday 7th to
  Friday 11th June 2021, Technical University of Denmark, Copenhagen Denmark\/}
  pp 511--512 \urlprefix\url{https://www.euspen.eu/knowledge-base/ICE21222.pdf}

\bibitem{Liu2020}
{Liu} Y, {Röse} A, {Prellinger} G, {Köchert} P, {Zhu} J and {Pollinger} F
  2020 {\em Journal of Lightwave Technology\/} {\bf 38} 1945--1952 ISSN
  1558-2213 \urlprefix\url{https://doi.org/10.1109/JLT.2019.2960473}

\bibitem{Sauthoff2023}
Sauthoff A, K\"{o}chert P, Prellinger G, Meyer T, Pilarski F, Weinrich S,
  Schmaljohann F, Guillory J, Truong D, Silbermann J, Kallio U, Jokela J and
  Pollinger F 2023 Two multi-wavelength interferometers for large-scale
  surveying {\em Proceedings of the 5th Joint International Symposium on
  Deformation Monitoring (JISDM)\/} (Valencia, Spain)
  \urlprefix\url{https://dx.doi.org/10.4995/JISDM2022.2022.13635}

\bibitem{Koechert2012}
K\"ochert P, Fl\"ugge J, Weichert C, K\"oning R and Manske E 2012 {\em
  Measurement Science and Technology\/} {\bf 23} 074005
  \urlprefix\url{https://doi.org/10.1088/0957-0233/23/7/074005}

\bibitem{Pollinger2026}
Pollinger F, Meyer T, Pilarski F, Yan H, Sauthoff A, Prellinger G, Weinrich S,
  Schmaljohann F and Köchert P 2026 {Research Data for the Publication
  'Unlimited-Range Absolute Distance Interferometry Outdoors via Tunable
  Synthetic and Pseudo-Supersynthetic Wavelengths: The TeleYAG-II
  Interferometer'} Physikalisch-Technische Bundesanstalt (PTB)
  \urlprefix\url{https://doi.org/10.7795/720.20261014}

\bibitem{Heydemann1981}
Heydemann P~L~M 1981 {\em Appl. Opt.\/} {\bf 20} 3382--3384
  \urlprefix\url{https://doi.org/10.1364/AO.20.003382}

\bibitem{Koening2015}
Köning R, Wimmer G and Witkovský V 2015 {\em Measurement Science and
  Technology\/} {\bf 26} 084004
  \urlprefix\url{https://doi.org/10.1088/0957-0233/26/8/084004}

\bibitem{Pollinger2020}
Pollinger F 2020 {\em Appl. Opt.\/} {\bf 59} 9771--9772
  \urlprefix\url{https://doi.org/10.1364/AO.400796}

\bibitem{Howe1981}
Howe D~A, Allan D~W and Barnes J~A 1981 Properties of singal sources and
  measurement methods {\em Proc. 35th Annual Symp. on Frequency Control\/} pp
  14--60 \urlprefix\url{https://tf.nist.gov/general/pdf/554.pdf}

\bibitem{Shmagun2022}
Shmagun V, Gerhardt U, Fröhlich T, Manske E and Kissinger T 2022 {\em
  Measurement Science and Technology\/} {\bf 33} 125024
  \urlprefix\url{https://doi.org/10.1088/1361-6501/ac970a}

\bibitem{Rustom2026}
Rustom Z, Ph\`{u}ng D~H, Chab\'{e} J, Courde C, M\'{e}tris G, Samain E, Raymond
  N, Scariot J, Mariey H, Maurice N and Artaud G 2026 {\em Opt. Express\/} {\bf
  34} 12269--12286 \urlprefix\url{https://doi.org/10.1364/OE.585413}

\bibitem{BIPM}
BIPM, IEC, IFCC, ILAC, ISO, IUPAC, IUPAP and OIML Evaluation of measurement
  data --- {G}uide to the expression of uncertainty in measurement Joint
  Committee for Guides in Metrology, JCGM 100:2008
  \urlprefix\url{https://www.bipm.org/documents/20126/2071204/JCGM\_100\_2008\_E.pdf/cb0ef43f-baa5-11cf-3f85-4dcd86f77bd6}

\bibitem{RbSpecs}
SI Scientific Instruments GmbH {\em Frequency Standards: PRS10 — Rubidium
  frequency standard with low phase noise\/}
  \urlprefix\url{https://www.si-gmbh.de/wp-content/uploads/2021/02/PRS10c_SI.pdf}

\bibitem{Andreas2018}
Andreas B and Weichert C 2018 Picometre level displacement interferometry {\em
  Modern Interferometry for Length Metrology\/} 2053-2563 (IOP Publishing) pp
  9--1 to 9--46 ISBN 978-0-7503-1578-4
  \urlprefix\url{https://doi.org/10.1088/2053-2563/aadddcch9}

\bibitem{Bender1965}
Bender P~L and Owens J~C 1965 {\em Journal of Geophysical Research\/} {\bf 70}
  2461--2462 ISSN 2156-2202
  \urlprefix\url{http://dx.doi.org/10.1029/JZ070i010p02461}

\bibitem{MeinersHagen2008}
Meiners-Hagen K and Abou-Zeid A 2008 {\em Measurement Science and Technology\/}
  {\bf 19} 084004 \urlprefix\url{https://doi.org/10.1088/0957-0233/19/8/084004}

\end{thebibliography}
\providecommand{\newblock}{}
%

\end{document}